\documentclass[prd,onecolumn,amsmath,amssymb,floatfix,superscriptaddress,nofootinbib]{revtex4-1}
\usepackage{bm}
\usepackage{amsmath}
\usepackage{epsfig}
\usepackage{xcolor}
\usepackage{natbib}
\usepackage{textcase}
\usepackage{graphicx}
\usepackage{ifthen}
\usepackage{xstring}
\usepackage{bbm}
\usepackage{graphicx}
\usepackage[utf8]{inputenc} 
\usepackage{amssymb}
\usepackage{csquotes}
\usepackage{latexsym}
\usepackage{epstopdf}
\usepackage{comment}
\usepackage{footmisc}

\DeclareGraphicsExtensions{.ps, .png}
\usepackage{dcolumn} 
\usepackage{multirow}
\usepackage{appendix}
\usepackage{footnote}
\usepackage{mathtools}
\usepackage{tabularx,ragged2e,booktabs}
\usepackage[normalem]{ulem}
\usepackage{float}
\restylefloat{table}

\definecolor{RedWine}{rgb}{0.743,0,0}
\definecolor{GrassGreen}{rgb}{0.125,0.75,0.125}
\definecolor{RoyalBlue}{rgb}{0.25,0.41,0.88}
\definecolor{darkgreen}{cmyk}{0.85,0.2,1.00,0.2} 
\definecolor{purple}{cmyk}{0.5,1.0,0,0} 
\definecolor{ultramarine}{rgb}{0.07, 0.04, 0.56}
\definecolor{cadmiumgreen}{rgb}{0.0, 0.42, 0.24}
\definecolor{indigo(dye)}{rgb}{0.0, 0.25, 0.42}

\newcommand{\Mpch}{\mbox{Mpc} h^{-1}}
\newcommand{\iMpch}{h \mbox{Mpc}^{-1}}
\newcommand{\fnl}{f_{\rm NL}}

\newcommand{\appropto}{\mathrel{\vcenter{
  \offinterlineskip\halign{\hfil$##$\cr
    \propto\cr\noalign{\kern2pt}\sim\cr\noalign{\kern-2pt}}}}}

\def\lsim{\mathrel{\raise.3ex\hbox{$$<$$\kern-.75em\lower1ex\hbox{$\sim$}}}}
\def\gsim{\mathrel{\raise.3ex\hbox{$$>$$\kern-.75em\lower1ex\hbox{$\sim$}}}}

\newcommand{\be}{\begin{equation}}
\newcommand{\ee}{\end{equation}}
\newcommand{\bea}{\begin{eqnarray}}

\def\({\left(}
\def\){\right)}
\def\<{\left\langle}
\def\>{\right\rangle}
\newcommand{\eea}{\end{eqnarray}}
\def\ba#1\ea{\begin{align}#1\end{align}}

\def\dd{\mathrm{d}}
\newcommand{\vs}{\nonumber\\} 
\makeatletter
\renewcommand*\l@section[2]
  {%
    \ifnum \c@tocdepth >\z@
      \addpenalty \@secpenalty
      \addvspace {1.0em \@plus \p@ }%
      \setlength \@tempdima {1.5em}%
      \begingroup
        \parindent \z@
        \rightskip \@pnumwidth
        \parfillskip -\@pnumwidth
        \leavevmode \bfseries
        \advance \leftskip \@tempdima
        \hskip -\leftskip
        #1\nobreak
        \hfil
        \nobreak
        \hb@xt@ \@pnumwidth {\hss #2\kern -\p@ \kern \p@ }%
        \par
      \endgroup
    \fi
  }
\makeatother

\def\vr{{\bm{r}}}
\def\vx{{\bm{x}}}

\def\vk{{\bm{k}}}

\def\vq{{\bm{q}}}

\def\Gaunt#1#2#3#4#5#6{\mathcal{G}^{#1}_{#4}{}^{#2}_{#5}{}^{#3}_{#6}}

\def\khat{{\hat{\bm{k}}}}

\def\qhat{{\hat{\bm{q}}}}
\def\rhat{{\hat{\bm{r}}}}

\def\fnl{f_\mathrm{NL}}

\def\obs{\mathrm{obs}}

\def\min{\mathrm{min}}

\def\dd{\mathrm{d}}

\def\myapp#1#2{%
  \mathrel{%
    \setbox0=\hbox{$#1\sim$}%
    \setbox2=\hbox{%
      \rlap{\hbox{$#1\propto$}}%
      \lower1.1\ht0\box0%
    }%
    \raise0.25\ht2\box2%
  }%
}

\usepackage[linktocpage=true]{hyperref}
\hypersetup{
colorlinks=true,
citecolor=ultramarine,
linkcolor=cadmiumgreen,
urlcolor=indigo(dye),
pdfauthor={},
pdftitle={},
pdfsubject={}
}

\usepackage[capitalise]{cleveref}

\begin{document}

\title{The galaxy bispectrum in the Spherical Fourier-Bessel  basis }

\author{Joshua N. Benabou}
\email{joshua\_benabou@berkeley.edu}

\affiliation{Berkeley Center for Theoretical Physics, University of California, Berkeley, CA 94720, U.S.A.}
\affiliation{Theoretical Physics Group, Lawrence Berkeley National Laboratory, Berkeley, CA 94720, U.S.A.}
\author{Adriano Testa}
\affiliation{California Institute of Technology, 1200 E. California Boulevard, Pasadena, CA 91125, USA}

\author{Chen Heinrich}
\affiliation{California Institute of Technology, 1200 E. California Boulevard, Pasadena, CA 91125, USA}

\author{Henry S. Grasshorn Gebhardt}
\affiliation{California Institute of Technology, 1200 E. California Boulevard, Pasadena, CA 91125, USA}
\affiliation{Jet Propulsion Laboratory, California Institute of Technology, Pasadena, California 91109, USA}

\author{Olivier Dor\'e}
\affiliation{California Institute of Technology, 1200 E. California Boulevard, Pasadena, CA 91125, USA}
\affiliation{Jet Propulsion Laboratory, California Institute of Technology, Pasadena, California 91109, USA}

\begin{abstract}
The bispectrum, the three-point correlation in Fourier space, is a crucial statistic for studying many effects targeted by the next-generation galaxy surveys, such as primordial non-Gaussianity (PNG) and general relativistic (GR) effects on large scales. In this work we develop a formalism for the bispectrum in the Spherical Fourier-Bessel (SFB) basis -- a natural basis for computing correlation functions on the curved sky, as it diagonalizes the Laplacian operator in spherical coordinates. Working in the SFB basis allows for line-of-sight effects such as redshift space distortions (RSD) and GR to be accounted for exactly, i.e without having to resort to perturbative expansions to go beyond the plane-parallel approximation. 
Only analytic results for the SFB bispectrum exist in the literature given the intensive computations needed. We numerically calculate the SFB bispectrum for the first time, enabled by a few techniques: We implement a template decomposition of the redshift-space kernel $Z_2$ into Legendre polynomials, and separately treat the PNG and velocity-divergence terms. We derive an identity to integrate a product of three spherical harmonics connected by a Dirac delta function as a simple sum, and use it to investigate the limit of a homogeneous and isotropic Universe. Moreover, we present a formalism for convolving the signal with separable window functions, and use a toy spherically symmetric window to demonstrate the computation and give insights into the properties of the observed bispectrum signal. While our implementation remains computationally challenging, it is a step toward a feasible full extraction of information on large scales via a SFB bispectrum analysis.

\end{abstract}
\maketitle

\onecolumngrid
\tableofcontents
\pagebreak
\section{Introduction}
\label{sec:intro}

Current and next-generation large-scale-structure (LSS) surveys such as DESI~\cite{DESI:2016fyo}, Euclid~\cite{Amendola:2016saw}, SPHEREx~\cite{dore2015cosmology} and the \textit{Nancy Grace Roman} Space Telescope~\cite{Eifler:2020hoy} will measure the galaxy density field over increasingly larger angular scales, enabling us to constrain interesting physical effects that become important on those scales, such as primordial non-Gaussianity~\cite{Dalal:2007cu} (PNG) and general relativistic effects~\cite{McDonald_2009}.

While many of our current techniques for estimators and modeling are well-suited for small-area surveys, they are challenged in larger surveys due to the breaking down of previously used approximations on the full sky. In particular, the plane-parallel approximation, which assumes that each galaxy has the same line-of-sight,  breaks down when the galaxy separation becomes large in a full-sky survey. Additionally, the Newtonian modeling of galaxy density also breaks down as general relativistic effects that grow as $1/k$ become important on large scales (for details see Refs.~\cite{Yoo_2012, Yoo_2010, Bonvin_2011, Challinor_2011}).

More precisely, redshift space distortions (RSD) induce effects in the observed galaxy density field that depend on the line-of-sight (LOS) of individual galaxies. Estimators assuming a fixed LOS for the entire survey will inevitably lose information at large galaxy separations. Even if one uses Yamamoto-like estimators~\cite{Yamamoto_2006, Bianchi_2015, Scoccimarro_2015} which assume a fixed LOS for each galaxy pair or triplet, there is still loss of information as the galaxies could have large angular separations in a given pair or triplet. The signal picked up by the Yamamoto estimator also includes wide-angle effects that are usually modeled either perturbatively as an expansion in the angular separation of the galaxy pair (i.e an expansion whose zero-order term is the plane-parallel approximation) \cite{Reimberg_2016,Beutler+:2019JCAP...03..040B,inpress,pardede2023wideangle}, or non-perturbatively via an exact calculation in the correlation function space \cite{Dio_2016}.

This raises the question of whether the Fourier basis is the optimal basis to use on the full sky. Indeed, the Fourier basis consists of the eigenfunctions of the Laplacian in Cartesian coordinates; the spherical Fourier-Bessel (SFB) basis, consisting of the eigenfunctions of the Laplacian in spherical coordinates, is a more natural basis for data analysis on the curved sky. The SFB basis was proposed for studying galaxy surveys since the early 90's~\cite{lahav1993spherical,Binney1991GaussianRF}, and was applied to data in the context of a power spectrum analysis in Refs.~\cite{Heavens_1995,Tadros_1999, Percival_2004}. 

Recently, an important limitation of the SFB analysis has been overcome in Ref.~\cite{samushia2019proper}, rendering computations of the power spectrum much more feasible: a boundary condition at the lower end of the redshift range was introduced, avoiding the need to carry many modes to model vanishing power outside of the survey footprint, which can introduce numerical instabilities. Later, the authors of Ref.~\cite{gebhardt2021superfab} developed a SFB power spectrum estimator with a public code release, which builds on this improvement as well as those in Ref.~\cite{Wang_2020} on pixel window effects and the separation of angular and radial transforms, making a SFB analysis feasible for surveys measuring the power spectrum such as \textit{Nancy Grace Roman}, SPHEREx and Euclid (see also \cite{Gebhardt:2021tme, Gebhardt:2023kfu}).

An alternative to the SFB basis called tomographic spherical harmonics (TSH) has also been explored in the literature, where the galaxy density contrast in a redshift bin is decomposed into spherical harmonics, and many redshift bins are used. In the limit of thin bins, neighboring bins are highly correlated, and the covariance matrix could become nearly degenerate. For thick bins, one loses information about the radial modes that are smaller than the bin size. SFB modes, in contrast, are more efficient basis functions since the radial modes are captured by spherical Bessel functions which are orthogonal to each other, unlike in the case of the redshift bin decomposition. See Ref.~\cite{Zhang_2021} for a detailed analysis comparing the SFB and TSH power spectrum (at $\Delta z = 0.1$) for current and future surveys, showing better $f_{\mathrm{NL}}$ constraints in general for the SFB method.

Limited effort, however, has been dedicated to the study of the SFB bispectrum. The bispectrum is the $3$-point correlation function in Fourier space and is of great importance to next-generation surveys. It is shown to be powerful at breaking parameter degeneracies when combined with the power spectrum for constraining galaxy bias parameters, neutrino masses and primordial non-Gaussianities (see e.g.~\cite{Gualdi_2020,Heinrich:2023qaa, Coulton_2019, Ruggeri_2018, dore2015cosmology}); the odd-parity bispectrum is also a smoking-gun signature for general relativistic effects that become more important on large scales~\cite{Maartens_2020, Jeong:2019igb, Clarkson:2018dwn}. 

A comprehensive derivation of the SFB bispectrum including all first and second-order GR effects, geometric effects and PNG was achieved in Ref.~\cite{bertacca}. However, due to the complexity of the computations involved, there has not yet been any work numerically evaluating the SFB bispectrum signal. In fact, while most of the integrals involved in calculating the signal are three-dimensional and are doable, some of the most important and interesting ones involving RSD and PNG contributions are four-dimensional and are intractable to compute naively.

In this paper, we derive a mathematical identity to express the six dimensional angular integral of three spherical harmonics connected by a Dirac-delta function as a simple sum, and use it to study the bispectrum signal in a homogeneous and isotropic Universe, to build intuition for later understanding the observed bispectrum in a realistic Universe. We also use this identity to accelerate the computation of RSD and PNG terms contributing to the observed bispectrum signal. Furthermore, we employ a template decomposition of the second-order coupling kernel in redshift space $Z_2$ into products of Legendre polynomials to evaluate all three-dimensional integrals. These techniques allow us to calculate and visualize for the first time the SFB bispectrum signal. 

We apply a general formalism we develop for convolution with a separable window function (in the angular and radial direction) to the toy example of a spherically symmetric window function to obtain numerical results that we study in detail. We derive key insights into the properties of the observed SFB bispectrum in a realistic Universe, highlighting those due to geometric effects.

The structure of this paper is as follows. In Section \ref{sec:background} we review the SFB basis and the modeling of the SFB galaxy power spectrum; we also describe the modeling of the Fourier space tree-level galaxy bispectrum and define the SFB bispectrum. Then in Section~\ref{sec:isohomo} we explore the calculation of the SFB bispectrum in the simplest case of a homogeneous and isotropic Universe, building up key intuition for interpreting the features of the observed bispectrum in the next section. In 
Section~\ref{sec:observed_sfb_s}, we incorporate various observational effects into the bispectrum, including growth of structure, galaxy bias, RSD, PNG and the survey window function. We present our template decomposition technique to enable its calculation, deferring the details of the derivation to the appendices; we then visualize and analyze the behavior of the observed bispectrum signal. Finally, we conclude and discuss future work in Section~\ref{sec:conclusion}. 

\section{Background}
\label{sec:background}

In this section, we begin by reviewing the SFB formalism and the SFB power spectrum following Ref.~\cite{gebhardt2021superfab}, to which we refer the readers for more details. We then review the Fourier space bispectrum, and describe the modeling of the observed galaxy bispectrum in redshift space including local PNG. Finally, we define the SFB bispectrum, which we later compute in Sections~\ref{sec:isohomo} and~\ref{sec:observed_sfb_s}. 

\subsection{The SFB formalism}

The spherical Fourier-Bessel mode  $\delta_{\ell m}(k)$ of the 
density contrast field $\delta(\vr)$ is defined by
\ba
\delta_{\ell m}(k) &\equiv
\int\dd^3\vr
\left[\sqrt{\frac{2}{\pi}}\,k\,j_\ell(kr)\,Y^*_{\ell m}(\rhat)\right]
\delta(\vr)\,.
\label{eq:sfb_fourier_pair_b}
\ea
The inverse transform is then 
\ba
\delta(\vr) &= \int\dd k\,\sum_{\ell m}
\left[\sqrt{\frac{2}{\pi}}\,k\,j_\ell(kr)\,Y_{\ell m}(\rhat)\right]
\delta_{\ell m}(k)\,,
\label{eq:sfb_fourier_pair_a}
\ea
where $\vr=r\rhat$ is the position vector, $r$ is the comoving
distance from the origin, and $\rhat$ is the line-of-sight direction.  

The spherical Fourier-Bessel modes are related to the Fourier modes via
\ba
\delta_{\ell m}(k)=\frac{k}{(2\pi)^\frac32}
\,i^{\ell}
\int\dd^2\khat
\,Y^*_{\ell m}(\khat)
\,\delta(\vk)
\,,
\label{eq:deltak2deltaklm}
\ea
for which the inverse relation is
\begin{equation}
\delta(\vk)
= \frac{(2\pi)^\frac32}{k}
\sum_{\ell m}
i^{-\ell}
\,Y_{\ell m}(\khat)
\,\delta_{\ell m}(k)
\label{eq:delta_from_deltalm}\,.
\end{equation}

Note that in this paper we use the following convention for the Fourier transform: 
\ba
f(\vk) &= \int\dd^3r\,e^{-i\vk\cdot\vr}\,f(\vr)\,,
\label{eq:fourier_convention_eq1}
\\
f(\vr) &= \int\frac{\dd^3k}{(2\pi)^3}\,e^{i\vk\cdot\vr}\,f(\vk)\,.
\label{eq:fourier_convention_eq2}
\ea
If unambiguous, we use the same symbol in configuration space (e.g, $f(\vr)$) as in Fourier space (e.g, $f(\vk)$). 

\subsection{Observed galaxy SFB power spectrum}

Let us denote the expansion of the matter density in cosmological perturbation theory by 
\begin{equation}
\delta(\vk,r) = \sum_{n=1}^{\infty} D^n(r)
\delta^{(n)}(\vk) \,,
\label{eq:ptansatz}
\end{equation}
with $D(r)$ the growth factor. Then, in the linear regime, the observed galaxy density contrast to first order can be modeled as
\ba
\delta^{g,\obs,(1)}(\vr)
&=
W(\vr)\,D(r)
\int\frac{\dd^3\vq}{(2\pi)^3}\,e^{i\vq\cdot\vr}
\,\widetilde A_\mathrm{RSD}(\mu,q\mu,r)b(r,q)
\,\delta^{(1)}(\vq)
\,,
\label{eq:observed_galaxy_density_linear}
\ea
where $W(\vr)$ is the survey window, $b(r,q)$ is
the linear galaxy bias, $\mu=\rhat\cdot \qhat$, and $\delta^{(1)}(\vq)$ is the matter density contrast in Fourier space. In what follows, the matter density contrast will always be denoted $\delta$ without any superscript, while the galaxy density contrast is denoted by $\delta^{g}$.

RSD effects are contained in $\widetilde A_\mathrm{RSD}(\mu,q\mu,r)$, which can be modeled as
\ba
\label{eq:Arsd}
\widetilde A_\mathrm{RSD}(\mu,q\mu,r)= \(1+\beta\mu^2\) \widetilde A_\mathrm{FoG}(q\mu)\,,
\ea
where $\beta=f/b$ and $f=\dd\ln D/\dd\ln a$ (with $a$ the scale factor) is the linear growth rate. In this paper we ignore the Fingers-of-God effect and set $\widetilde A_\mathrm{FoG}(q\mu)=1$.

Transforming to spherical Fourier-Bessel space we have that
\ba
\delta^{g,\obs, (1)}_{\ell m}(k)
&=
\int\dd q
\sum_{LM}
\mathcal{W}_{\ell m}^{LM}(k,q)
\,\delta_{LM}^{(1)}(q)\,,
\label{eq:sfb_coeff_obs}
\ea
where the observed galaxy density $\delta^{g,\obs}_{\ell m}(k)$ is related to the matter density $\delta_{LM}(q)$ by the mode coupling matrix $\mathcal{W}_{\ell m}^{LM}(k,q)$. In our convention, this mode coupling matrix encodes galaxy physics such as galaxy bias and RSD effects, unequal time effects such as the growth of structure, and the survey window function $W(\mathbf{r})$:
\ba
\mathcal{W}_{\ell m}^{LM}(k,q)
&=
\int\dd^2\rhat
\,Y_{LM}(\rhat)
\,Y^*_{\ell m}(\rhat)
\,\mathcal{W}_\ell^L(k,q,\rhat)\,,
\label{eq:sfb_wlmLMkq}
\ea
where 
\ba
\mathcal{W}_\ell^L(k,q,\rhat)
&=
\frac{2qk}{\pi}
\int\dd r\,r^2
\,W(\vr) \,D(r)\,b(r,q)
\,j_\ell(kr)\widetilde A_\mathrm{RSD}(-i\partial_{qr},-iq\partial_{qr},r) j_{L}(qr)\, ,
\label{eq:sfb_wlLkqrhat}
\ea
where we replace the argument $\mu$ of $\widetilde{A}_{\rm RSD}$ by $-i\partial_{qr}$ which acts on $e^{i\vq\cdot\vr}=e^{iqr\mu}$.

Noting that in a homogeneous and isotropic Universe, the matter power spectrum satisfies 
\begin{equation}
\langle\delta(\boldsymbol{k}) \delta^{*}\!\!\left(\boldsymbol{k}^{\prime}\right)\rangle=(2 \pi)^{3} \delta^{D}\left(\boldsymbol{k}-\boldsymbol{k}^{\prime}\right) P(k) \,,
\end{equation}
it follows that the 2-point function of the SFB modes is
\ba
&\<\delta^{g,\obs}_{\ell m}(k)\,\delta^{g,\obs,*}_{\ell'm'}(k')\>= 
\int\dd q
\sum_{LM}
\mathcal{W}_{\ell m}^{LM}(k,q)
\,\mathcal{W}_{\ell'm'}^{LM,*}(k',q)
\,P(q)\,.
\label{eq:sfb_power_spectrum_window}
\ea

In the full-sky limit where $W(\mathbf{r})=W(r)$, we have that $\mathcal{W}^{L}_{\ell}(k,q,\rhat) \rightarrow \mathcal{W}^{\ell}_{\ell}(k,q)$ is independent of $\rhat$. Let us define $\mathcal{W}_{\ell}(k,q) \equiv \mathcal{W}^{\ell}_{\ell}(k,q)$. Then the SFB power spectrum $C_\ell(k,k')$ defined via
\ba
\<\delta^{g,\obs}_{\ell m}(k)\,\delta^{g,\obs,*}_{\ell'm'}(k')\>
&=
\delta^K_{\ell\ell'}
\delta^K_{mm'}
\,C_\ell(k,k')\,,
\label{eq:sfb_power_spectrum}
\ea
can be expressed as 
\ba
C_\ell(k,k')
&=
\int\dd q
\,\mathcal{W}_{\ell}(k,q)
\,\mathcal{W}^*_{\ell}(k',q)
\,P(q)\,, 
\label{eq:Cl}
\ea
where
\ba
\mathcal{W}_\ell(k,q)=
\frac{2qk}{\pi}
\int\dd r\,r^2
\,W(r) \,D(r)
\,j_\ell(kr)
\bigg(b(r,q)j_\ell(qr) - f(r) j_\ell^{''}(qr)\bigg),
\label{eq:W_lkq_first}
\ea
where we use Eq.~\ref{eq:Arsd} with $\mu\to-i\partial_{qr}$.

\begin{figure*}[ht]
\centering
\includegraphics[width=0.53\linewidth]{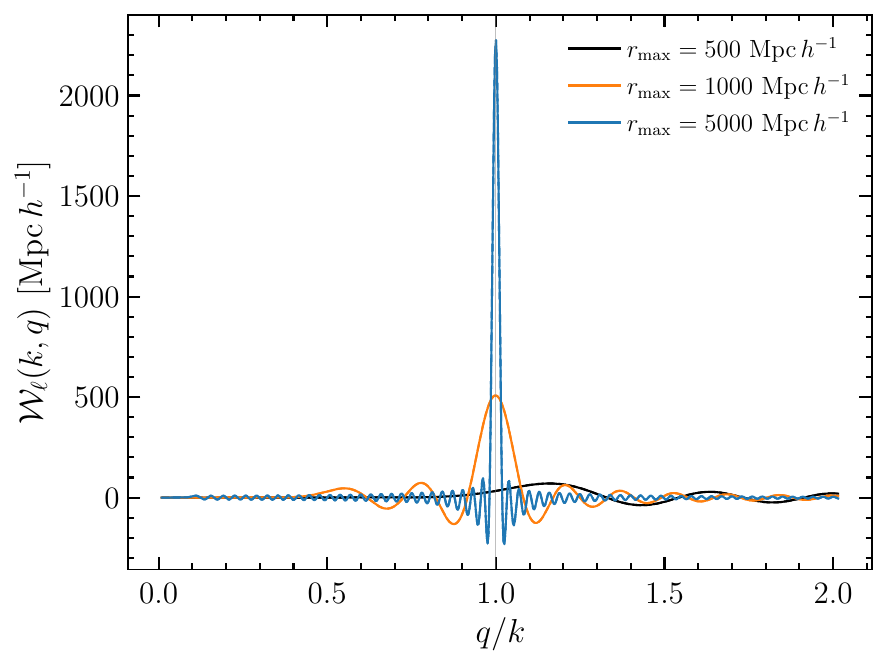}
\caption{$\mathcal{W}_\ell(k,q)$ for fixed $\ell=20$, $k=4.18 \times 10^{-2} \, \iMpch$ and
for various sizes $r_\mathrm{max}$ of the survey window $W(r)=\mathbf{1}_{[0,r_\mathrm{max}]}(r)$.
}
\label{fig:Wlkq}
\end{figure*}

Note that in a homogeneous and isotropic Universe, for which $b(r,q) = D(r) = \widetilde A_\mathrm{RSD} = W(\vr) = 1$, $\mathcal{W}_{\ell}(k,q)$ becomes a Dirac delta function and we have that $C_l(k,k') = \delta_D(k-k') P(k)$. In reality, the kernels $\mathcal{W}_\ell(k,q)$ are peaked at $k \approx q$. We show examples of $\mathcal{W}_\ell(k,q)$ for the spherical window $W(r)=\mathbf{1}_{[0,r_\mathrm{max}]}(r)$ for various values of $r_{\rm max}$ in Fig.~\ref{fig:Wlkq}, where we fix $k=4.18 \times 10^{-2}$ and vary $q/k$ for $\ell=20$. Here and in the remainder of this work we use the Planck 2018 cosmology \cite{Aghanim:2018eyx} as our fiducial cosmology. The matter power spectrum at zero redshift and the linear growth factors $f$ and $D$ are computed with the Boltzmann code \texttt{CAMB}\footnote{\url{https://camb.info/}}~\cite{camb}. All other calculations are performed in Julia \cite{Julia-2017}.

\subsection{Observed Fourier galaxy bispectrum}
\label{sec:SFBbispectrum_def}
We now review the observed galaxy bispectrum in Fourier space including observational effects such as the RSD and galaxy bias, but without the window function convolution. For details on the derivation of the various quantities, we refer the readers to Ref.~\cite{Tellarini_2016} and~\cite{Bernardeau_2002} which we follow closely. As we will also be concerned with modeling the effects of PNG in the SFB bispectrum, we will include its effects in the Fourier bispectrum as well. 

We consider PNG of the local type, for which the fluctuations of the potential are parameterized by
\ba
\Phi_{\mathrm{NG}}(\boldsymbol{x})=\varphi(\boldsymbol{x})+f_{\mathrm{NL}}\left(\varphi^{2}(\boldsymbol{x})-\left\langle\varphi^{2}\right\rangle\right) \,,
\ea
where $\varphi$ is a primordial Gaussian potential.
Using the Poisson equation, we may relate the long-wavelength Gaussian potential to the linearly evolved primordial matter density perturbation via
\ba
\Phi_{\mathrm{NG}}(\boldsymbol{k})=\frac{\delta^{(1)}(\boldsymbol{k}, z)}{\alpha(k, z)} \,,
\label{eq:Phi_NG}
\ea
 where
\ba
\alpha(k, z)=\frac{2 k^{2} c^{2} D(z) T(k)}{3 H_{0}^{2} \Omega_{\mathrm{m}}}\,.
\label{eq:alpha_def}
\ea
Here $\Omega_m$ is the matter density, $H_0$ is the Hubble constant, and $T(k)$ is the transfer function of matter perturbations, normalized to $1$ at low $k$. Eq.~\ref{eq:Phi_NG}-\ref{eq:alpha_def} are valid in the Newtonian limit on subhorizon scales \cite{Wands:2009ex, Yoo:2009au, Yoo:2010ni}. 

In perturbation theory, the observed galaxy density contrast field at position $\vr$ is given by
\begin{eqnarray}
\delta^{g}(\vk,\vr) &=& \sum_{n=1}^{\infty} D^n(r) \int \frac{\dd^3\vk_1}{(2\pi)^3}
\ldots \int \frac{\dd^3\vk_n}{(2\pi)^3} \,(2\pi)^3\delta_D(\vk_1+\dots+\vk_n-\vk)\, Z_n (\vk_1, \ldots, \vk_n,\vr) \,
\delta^{(1)}(\vk_1) \cdots \delta^{(1)}(\vk_n)\,,
\label{eq:Zn}
\end{eqnarray}
where $\delta^{(1)}$ is the linear matter density field, and the $n$-th order redshift space kernels $Z_{n}$ encode the mode coupling effects from gravitational evolution, PNG and galaxy biasing. We assume the  bivariate galaxy biasing model 
\begin{equation}
\begin{aligned}
\delta^{\mathrm{g}}(\boldsymbol{x}) & =b^\mathrm{E}_{10} \delta(\boldsymbol{x})+b^\mathrm{E}_{01} \varphi(\boldsymbol{x})  + b^\mathrm{E}_{20}\left(\delta(\boldsymbol{x})\right)^2+b^\mathrm{E}_{11} \delta(\boldsymbol{x}) \varphi(\boldsymbol{x})  +b^\mathrm{E}_{02}(\varphi(\boldsymbol{x}))^2+ b_{s_2}\left(s^2-\left\langle s^2\right\rangle\right)-b^\mathrm{E}_{01} n^2 \,.
\end{aligned}
\end{equation}
Above we define the tidal term \cite{Catelan:2000vn, 2012PhRvD..86h3540B} 
\ba
s^2(\boldsymbol{k})=\int \frac{d \boldsymbol{q}}{(2 \pi)^3} S_2(\boldsymbol{q}, \boldsymbol{k}-\boldsymbol{q}) \delta^{(1)}(\boldsymbol{q}) \delta^{(1)}(\boldsymbol{k}-\boldsymbol{q}),
\ea
and the (non-Gaussian) term encoding displacement of galaxies from their initial Lagrangian coordinate positions $\boldsymbol{q}$
\ba
n^2(\boldsymbol{k})=2 \int \frac{d \boldsymbol{q}}{(2 \pi)^3} N_2(\boldsymbol{q}, \boldsymbol{k}-\boldsymbol{q}) \frac{\delta^{(1)}(\boldsymbol{q}) \delta^{(1)}(\boldsymbol{k}-\boldsymbol{q})}{\alpha(|\boldsymbol{k}-\boldsymbol{q}|)} \,,
\ea
where above we use the kernels 
\ba
N_{2}\left(\vk_{1}, \vk_{2}\right)&=\frac{\vk_{1} \cdot \vk_{2}}{2 k_{1}^{2}} \,, \\
S_{2}\left(\vk_{1}, \vk_{2}\right)&=\frac{\left(\vk_{1} \cdot \vk_{2}\right)^{2}}{k_{1}^{2} k_{2}^{2}}-\frac{1}{3}\,.
\ea

The redshift space kernels at first and second order, respectively, are given by
\ba
Z_{1}(\vk,\vr)&=b^\mathrm{E}_{10}+f(r)\mu^{2}+\frac{b^\mathrm{E}_{01}}{\alpha(k)}, \label{eq:z1}\\
Z_{2} (\vk_1,\vk_2,\vr)&=b^\mathrm{E}_{10}\left[F_{2}\left(\vk_{1}, \vk_{2}\right)+f_{\mathrm{NL}} \frac{\alpha(k)}{\alpha\left(k_{1}\right) \alpha\left(k_{2}\right)}\right]+\left[b^\mathrm{E}_{20}-\frac{2}{7} b_{10}^{\mathrm{L}} S_{2}\left(\vk_{1}, \vk_{2}\right)\right] \nonumber \\
&+\frac{b^\mathrm{E}_{11}}{2}\left[\frac{1}{\alpha\left(k_{1}\right)}+\frac{1}{\alpha\left(k_{2}\right)}\right]+\frac{b^\mathrm{E}_{02}}{\alpha\left(k_{1}\right) \alpha\left(k_{2}\right)}-b^\mathrm{E}_{01}\left[\frac{N_{2}\left(\vk_{1}, \vk_{2}\right)}{\alpha\left(k_{2}\right)}+\frac{N_{2}\left(\vk_{2}, \vk_{1}\right)}{\alpha\left(k_{1}\right)}\right] \nonumber \\
&+f(r) \mu^{2}\left[G_{2}\left(\vk_{1}, \vk_{2}\right)+f_{\mathrm{NL}} \frac{\alpha(k)}{\alpha\left(k_{1}\right) \alpha\left(k_{2}\right)}\right]+\frac{f(r)^{2} k^{2} \mu^{2}}{2} \frac{\mu_{1} \mu_{2}}{k_{1} k_{2}}+b^\mathrm{E}_{10} \frac{f(r) \mu k}{2}\left(\frac{\mu_{1}}{k_{1}}+\frac{\mu_{2}}{k_{2}}\right) \nonumber \\
&+b^\mathrm{E}_{01} \frac{f(r) \mu k}{2}\left[\frac{\mu_{1}}{k_{1} \alpha\left(k_{2}\right)}+\frac{\mu_{2}}{k_{2} \alpha\left(k_{1}\right)}\right],
\label{eq:z2_with_fnl}
\ea
where $\mu \equiv \khat \cdot {\rhat}$, $\vk \equiv
\vk_1 +\vk_2$, $\mu_i \equiv \khat_i \cdot {\rhat }$, and where the coupling kernels for the real-space density and velocity-divergence fields are
\ba
F_2(\vq_1, \vq_2)  &= \frac{5}{7} + \frac{1}{2} \bigg( \frac{q_1}{q_2} + \frac{q_2}{q_1} \bigg) \frac{\vq_1 \cdot \vq_2}{q_1 q_2 } + \frac{2}{7}\frac{(\vq_1 \cdot \vq_2)^2}{(q_1 q_2)^2}\,,
\label{eq:F2_appendix}\\
G_2(\vq_1,\vq_2) &= \frac{3}{7} + \frac{1}{2} \bigg(\frac{q_1}{q_2} + \frac{q_2}{q_1}\bigg)\frac{\vq_1 \cdot
\vq_2}{q_1 q_2}  + \frac{4}{7}
\frac{(\vq_1 \cdot \vq_2)^2}{(q_1 q_2)^2} \label{eq:G2}  \,.
\ea
From Eq.~\ref{eq:z1} it follows that we may write the linear galaxy bias appearing in Eq.~\ref{eq:W_lkq_first} as $b(r,q)=b^\mathrm{E}_{10} + b^\mathrm{E}_{01}/\alpha(q,r)$.

The galaxy bispectrum in Fourier space is defined via
\ba
\<\delta^{g}(\vk_1,\vr_1)\delta^{g}(\vk_2,\vr_2)\delta^{g}(\vk_3,\vr_3)\>=B_s(\vk_1,\vk_2,\vk_3,\vr_1,\vr_2,\vr_3) (2\pi)^3\delta_D(\vk_1+\vk_2+\vk_3) \,.
\label{eq:3point_bispectrum_fourier_final}
\ea
Working up to second order in the galaxy density field expansion, the tree-level bispectrum is
\ba
B_s(\vk_1,\vk_2,\vk_3,\vr_1,\vr_2,\vr_3) \equiv 2D(r_1)D(r_2)D^2(r_3)P(k_1)P(k_2)Z_1(\vk_1,\vr_1)Z_1(\vk_2,\vr_2)Z_2(\vk_1,\vk_2,\vr_3)+\mathrm{2\ cyc. \ perm.} \,,
\label{eq:Bs}
\ea
where we sum over all cyclic permutations of the subscripts of the quantities in parentheses.
Note that in the absence of RSD, linear growth, galaxy bias, and PNG, Eq.~\ref{eq:Bs} reduces to the matter bispectrum
\ba
B_m(\vk_1, \vk_2, \vk_3) \equiv 2P(k_1)P(k_2) F_2(\vk_1, \vk_2) 
+ \mathrm{2\ cyc. \ perm.}\,  
\,.
\label{eq:Bm}
\ea

We follow Ref.~\cite{Tellarini_2016} to model the Eulerian biases
\ba
&b_{10}^{\mathrm{E}}=1+b_{10}^{\mathrm{L}} \,, \\
&b_{01}^{\mathrm{E}}=b_{01}^{\mathrm{L}} \,,\\
&b_{20}^{\mathrm{E}}=\frac{8}{21} b_{10}^{\mathrm{L}}+b_{20}^{\mathrm{L}}\,, \\
&b_{11}^{\mathrm{E}}=b_{01}^{\mathrm{L}}+b_{11}^{\mathrm{L}} \,,\\
&b_{02}^{\mathrm{E}}=b_{02}^{\mathrm{L}} \,,
\intertext{in terms of the Lagrangian biases, which are given by}
b_{01}^{\mathrm{L}} &=2 f_{\mathrm{NL}} \delta_{c} b_{10}^{\mathrm{L}}, \\
b_{11}^{\mathrm{L}} &=2 f_{\mathrm{NL}}\left(\delta_{c} b_{20}^{\mathrm{L}}-b_{10}^{\mathrm{L}}\right), \\
b_{02}^{\mathrm{L}} &=4 f_{\mathrm{NL}}^{2} \delta_{c}\left(\delta_{c} b_{20}^{\mathrm{L}}-2 b_{10}^{\mathrm{L}}\right) \,,
\ea
if one assumes a Universal Mass Function, and where $\delta_c$ is the critical overdensity for halo collapse, here set to its value for spherical collapse $\delta_c = 1.686$.

Note that only $b_{10}^E$ and $b_{20}^E$ need to be specified in order to determine all the other bias parameters. Specifically for our SFB bispectrum computation later, we will set $b_{10}^E = 1.8$ and $b_{20}^E = 0.305$.  While there is no technical obstacle to including redshift-dependent biases in the SFB calculation, we choose flat biases here for simplicity.

\subsection{SFB bispectrum definition}
We now review the formalism for the SFB bispectrum. We seek to compute the $3$-point correlation function of the observed galaxy over-density field in SFB space:
\ba
\<\delta^{g,\obs}_{l_1 m_1}(k_1)\,\delta^{g,\obs}_{l_2 m_2}(k_2)\delta^{g,\obs}_{l_3 m_3}(k_3)\> \,.
\label{eq:sfbdef}
\ea
In the following, it will be useful to distinguish between two notions of isotropy, which we term \emph{observational} isotropy and \emph{intrinsic} isotropy. Intrinsic isotropy refers to the statistically isotropic distribution of galaxies on the largest-scales in real-space. Due to RSD, the galaxy clustering observed in surveys is not intrinsically isotropic since, in redshift space, it depends on the angle to a given LOS. On the other hand, the distribution observed by a full-sky survey remains invariant under rotations about the observer position. This observational isotropy is only broken by a survey window which is not  spherically symmetric. We show in Appendix \ref{app:isotropy} that, assuming observational isotropy, Eq.~\ref{eq:sfbdef} is real and proportional to the Gaunt factor 
\ba
\mathcal{G}^{l_1l_2l_3}_{m_1m_2m_3}
&\equiv
\int\dd^2\rhat
\,Y_{l_1 m_1}(\rhat)
\,Y_{l_2m_2}(\rhat)
\,Y_{l_3m_3}(\rhat)\,,
\label{eq:gaunt_definition}
\ea
which can be expressed in terms of Wigner-$3j$ symbols,
\ba
\mathcal{G}^{l_1l_2l_3}_{m_1m_2m_3}
&=
\(\frac{(2l_1+1)(2l_2+1)(2l_3+1)}{4\pi}\)^\frac12
\begin{pmatrix}
  l_1 & l_2 & l_3 \\
  0 & 0 & 0
\end{pmatrix}
\begin{pmatrix}
  l_1 & l_2 & l_3 \\
  m_1 & m_2 & m_3
\end{pmatrix}\,.
\ea
The Wigner-$3j$'s ensure the SFB $3$-point function vanishes unless the following conditions are satisfied: (i) $m_1+m_2+m_3=0$, (ii) triangle inequality on the $l_i$: $l_i \ge l_j-l_k$, and (iii) $l_1+l_2+l_3$ is even.

In order to rid of the purely geometric information contained in the $m_i$, we compute the ``angle-averaged" bispectrum 
\ba
B^{\text{SFB}}_{l_1l_2l_3}(k_1,k_2,k_3)&\equiv\sum_{m_1,m_2,m_3}\begin{pmatrix}
  l_1 & l_2 &  l_3 \\
  m_1 & m_2 & m_3
\end{pmatrix}\<\delta^{g,\obs}_{l_1m_1}(k_1)\delta^{g,\obs}_{l_2m_2}(k_2)\delta^{g,\obs}_{l_3m_3}(k_3)\> \,.
\label{eq:sfb_3point_reduced}
\ea
Using the orthogonality relation in Eq.~\ref{eq:gaunt_orthogonality} then gives
\ba
\<\delta^{g,\obs}_{l_1m_1}(k_1)\delta^{g,\obs}_{l_2m_2}(k_2)\delta^{g,\obs}_{l_3m_3}(k_3)\> =\begin{pmatrix}
  l_1 & l_2 &  l_3 \\
  m_1 & m_2 & m_3
\end{pmatrix}B^{\text{SFB}}_{l_1l_2l_3}(k_1,k_2,k_3) \,.
\label{eq:angle_averaged_bispectrum}
\ea
In the following subsections we will always plot the dimensionless reduced bispectrum 
\ba
Q^{\text{SFB}}_{l_1l_2l_3}(k_1,k_2,k_3) \equiv \frac{B^{\text{SFB}}_{l_1l_2l_3}(k_1,k_2,k_3)}{P(k_1)P(k_2) + P(k_1)P(k_3) + P(k_2)P(k_3) } \,,
\label{eq:reduced_SFB_bispec}
\ea
which partially projects out the dependence of the signal on $k_i$ coming from the matter power spectrum. Finally, we note that the bispectrum is invariant under simultaneous cyclic permutations of $(l_1,l_2,l_3)$ and $(k_1,k_2,k_3)$, which allows us to restrict to $l_1 \le l_2 \le l_3$.

Before delving into the computation of the SFB bispectrum, let us briefly remark on its relation to the angular bispectrum in spherical shells 
(i.e, the TSH bispectrum) 
$b_{l_1l_2l_3}(r_1,r_2,r_3)$ of Ref.~\cite{Dio_2016}, defined via
\ba
 \mathcal{G}^{l_1l_2l_3}_{m_1m_2m_3} b_{l_1l_2l_3}(r_1, r_2, r_3) \equiv \int d^2\rhat_1d^2 \rhat_2d^2\rhat_3Y^*_{l_1,m_1}(\rhat_1)Y^*_{l_2,m_2}(\rhat_2)Y^*_{l_3,m_3}(\rhat_3)\<\delta^{g, \obs}(\vr_1)\delta^{g, \obs}(\vr_2)\delta^{g, \obs}(\vr_3)\>\,.
\label{eq:dedio_bispec_def}
\ea
Using Eq.~\ref{eq:angle_averaged_bispectrum} in combination with Eq.~\ref{eq:sfb_fourier_pair_b}, it follows that
\ba
B^{\text{SFB}}_{l_1l_2l_3}(k_1,k_2,k_3) 
&=\bigg(\frac{2}{\pi}\bigg)^{\frac{3}{2}}k_1k_2k_3\int \left(\prod_i \dd r_i r_i^2 j_{l_i}(k_i r_i)\right)b_{l_1 l_2 l_3}(r_1, r_2, r_3)\,. \label{eq:SFB_bispectrum_observed}
\ea
We see that the SFB bispectrum and the TSH bispectrum are related by an invertible linear transformation, and that the multipole indices $l_i$ are the same in the SFB and TSH formalisms (the wavenumbers $k_i$ are the same between SFB and Fourier space).
In practice, in the TSH formalism many radial bins are desirable to fully exploit the large scale radial modes (see \cite{Zhang_2021}, which studied this for the power spectrum), leading to a covariance matrix which is difficult to invert, whereas this issue does not arise in the SFB basis.

\section{SFB bispectrum in a homogeneous and isotropic Universe}
\label{sec:isohomo}

\subsection{Key identity for fast computation}

We first examine the bispectrum in the limit of a homogeneous and intrinsically isotropic Universe (by ignoring the growth of structure, galaxy bias evolution, redshift-space distortions and window function effects) in order to study its features and build up intuition for understanding the observed SFB bispectrum later in Section~\ref{sec:observed_sfb_s}.

We can relate the SFB bispectrum to the Fourier bispectrum using the relation between the SFB and Fourier modes in Eq.~\ref{eq:deltak2deltaklm}: 
\ba
\<\delta_{l_1 m_1}(k_1)\,\delta_{l_2 m_2}(k_2)\delta_{l_3 m_3}(k_3)\>=\frac{k_1k_2k_3}{(2\pi)^{\frac{9}{2}}}i^{l_1+l_2+l_3}\int d^2\khat_1d^2 \khat_2d^2\khat_3Y^*_{l_1,m_1}(\khat_1)Y^*_{l_2,m_2}(\khat_2)Y^*_{l_3,m_3}(\khat_3)\<\delta(\vk_1)\,\delta(\vk_2)\delta(\vk_3)\> \,.
\label{eq:sfb_3point_expanded}
\ea
Due to homogeneity and isotropy, the Fourier bispectrum $B_{m}(k_1, k_2, k_3)$ (Eq.~\ref{eq:Bm}) depends only on the lengths $k_1,k_2$ and $k_3$, so that we may write 
\ba
\<\delta_{l_1 m_1}(k_1)\,\delta_{l_2 m_2}(k_2)\delta_{l_3 m_3}(k_3)\>&=\frac{k_1k_2k_3}{(2\pi)^{\frac{9}{2}}}i^{l_1+l_2+l_3}(2\pi)^3B_{m}(k_1, k_2, k_3) \; 
I_{m_1m_2m_3}^{l_1l_2l_3}(k_1,k_2,k_3),
\label{eq:sfb_3point_expanded1}
\ea
where
\begin{equation}
I_{m_1m_2m_3}^{l_1l_2l_3}(k_1,k_2,k_3) \equiv
\int d^2\khat_1d^2 \khat_2d^2\khat_3Y^*_{l_1,m_1}(\khat_1)Y^*_{l_2,m_2}(\khat_2)Y^*_{l_3,m_3}(\khat_3)\delta_D(\vk_1+\vk_2+\vk_3). 
\label{eq:I}
\end{equation}

Eq.~\ref{eq:I} has typically been written in terms of an integral over the spherical Bessel functions \cite{bertacca} (for details, see appendix \ref{app:derivation_identity}) as
\ba
I_{m_1m_2m_3}^{l_1l_2l_3}(k_1,k_2,k_3)=8i^{l_1+l_2+l_3}\int r^2\dd r j_{l_1}(k_1r)j_{l_2}(k_2r)j_{l_3}(k_3r)\int d^2\rhat Y^*_{l_1,m_1}(\rhat)Y^*_{l_2,m_2}(\rhat)Y^*_{l_3,m_3}(\rhat) \,.
\label{eq:gauntlike_propto_gaunt_main_text}
\ea
where the second integral is the Gaunt factor (Eq.~\ref{eq:gaunt_definition}). Here we instead derive the identity (see derivation in appendix \ref{app:derivation_identity})
\ba
I_{m_1m_2m_3}^{l_1l_2l_3}(k_1,k_2,k_3) 
&=\frac{4\pi^{\frac{3}{2}}}{k_1k_2k_3}\sqrt{(2l_1+1)}
\begin{pmatrix}
  l_1 & l_2 &  l_3 \\
  m_1 & m_2 & m_3
\end{pmatrix}  \sum_{|m| \le \min(l_2,l_3)} \!\!\!\!\!{Y_{l_2,m}}(\theta_{12},0)Y_{l_3,-m}(\theta_{13},0)
(-1)^m
\begin{pmatrix}
  l_1 & l_2 & l_3 \\
  0 & m & -m
\end{pmatrix}\,,
\label{eq:gauntlike_identity}
\ea
which allows us to rapidly compute the angle-averaged bispectrum without any numerical integration\footnote{For this purpose we precompute a lookup table of $Y_{lm}(\theta,0)$ values and interpolate. Also note that we may halve the number of terms in the sum by using its invariance under $m \to -m$.}
\ba
B^{\text{SFB, iso/homo}}_{l_1l_2l_3}(k_1,k_2,k_3)
&=
B_{m}(k_1, k_2, k_3) i^{l_1+l_2+l_3}
\sqrt{2(2l_1+1)} \nonumber \\ 
&\times
\sum_{|m| \le\min(l_2,l_3)}Y_{l_2,m}(\theta_{12},0)Y_{l_3,-m}(\theta_{13},0)
(-1)^m\begin{pmatrix}
  l_1 & l_2 & l_3 \\
  0 & m & -m
\end{pmatrix}  \,.
\label{eq:sfb_3point_expanded2}
\ea
Above we define $\theta_{ij}$ as the angle between $\vk_i$ and $\vk_j$, such that $\cos(\theta_{12}) \equiv \khat_1 \cdot \khat_2 = \vartheta(k_1,k_2,k_3)$, where $\vartheta(k_1,k_2,k_3) \equiv \frac{k_3^2-k_1^2-k_2^2}{2k_1k_2}$, and $\cos(\theta_{13}) \equiv \vartheta(k_1,k_3,k_2)$.
Furthermore, denoting the spherical coordinates of $\rhat$ by $(\theta,\phi)$, we define $Y_{\ell,m}(\rhat) \equiv Y_{\ell,m}(\theta,\phi)$.
It is clear from Eq.~\ref{eq:sfb_3point_expanded2} that the SFB bispectrum in an isotropic and homogeneous Universe is proportional to the Fourier bispectrum by a geometric coupling factor depending on the $l_i$ and the angles between the $\vk_i$. 
This factor also imposes the triangle inequality on the wavenumbers, i.e,  $k_i \le |k_j-k_k|$, which is relaxed as we shall see in the next section for the observed SFB bispectrum and is only imposed approximately for spherically symmetric surveys which extend to sufficiently large redshifts. 

The identity Eq.~\ref{eq:gauntlike_identity} is one of our key results.
In addition to trivializing the computation of the signal in a homogeneous and isotropic Universe, it provides analytic insight into the geometric features of the observed bispectrum.
Crucially, we will employ this identity to render the computation of the observed bispectrum tractable. We will discuss these points in \ref{sec:observed_sig_modeling_results}.

\begin{figure*}[!t]
\centering
\includegraphics[width=0.98\linewidth]{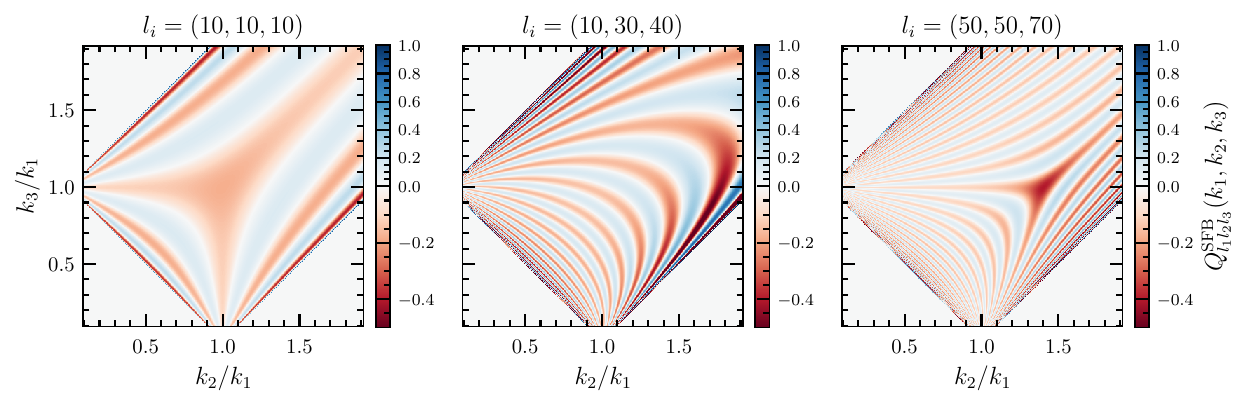} %
\caption{The reduced SFB bispectrum signal for an isotropic and homogeneous Universe, as a function of $k_2$ and $k_3$ in units of $k_1$, which is fixed to $k_1=4.18 \times 10^{-2} \ \iMpch$ here. Each panel displays a different triplet $l_i \equiv (l_1,l_2,l_3)$. The bispectrum vanishes identically for configurations $(k_1,k_2,k_3)$ which do not satisfy the triangle condition. The oscillations are a result of the geometric coupling in Eq.~\ref{eq:harmonic_product};  their number is controlled by the values of $l$. Note that the colorbar limits are saturated in each panel.
}
\label{fig:iso_homo_mixed_multipoles}
\end{figure*}
\subsection{Properties of the signal}
In Fig.~\ref{fig:iso_homo_mixed_multipoles}, we show two-dimensional cross-sections of the reduced bispectrum in an isotropic and homogeneous Universe as a function of $k_2/k_1$ and $k_3/k_1$ for fixed $k_1$, for three $l$-triplets $(l_1, l_2, l_3)$. The most striking feature is the rectangular border outside of which the signal vanishes; this is the enforcement of the triangle inequality.   

Another important feature is that the signal oscillates in the space of $k_i$'s, which is not surprising, given that the products of spherical harmonics in 
$I^{l_1l_2l_3}_{m_1 m_2 m_3}(k_1,k_2,k_3)$,
\ba
Y_{l_2,m}(\theta_{12},0)Y_{l_3,-m}(\theta_{13},0) \,,
\label{eq:harmonic_product}
\ea
oscillate as the angles between the $\vk_i$'s vary. Further, the number of oscillations as one moves from the center of the plot corresponding to an equilateral $k$-triangle, toward the borders of the rectangular region, corresponding to degenerate triangles, is higher for larger $l_i$ values.

\begin{figure*}[ht]
\centering
\includegraphics[width=0.53\linewidth]{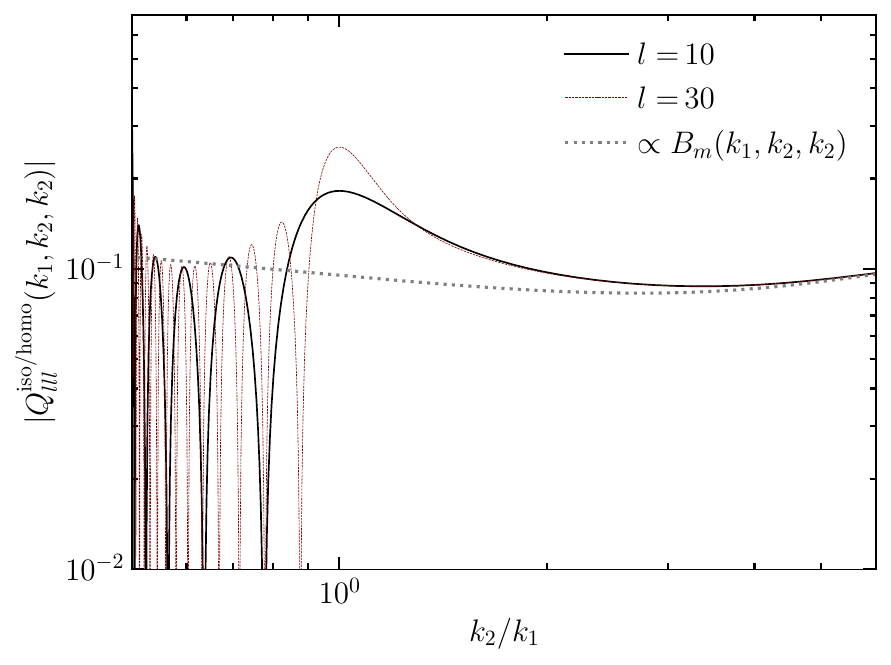}
\caption{The reduced bispectrum signal in an isotropic and homogeneous Universe for fixed $k_1=4.18 \times 10^{-2} \ \iMpch$ as in Fig.~\ref{fig:iso_homo_mixed_multipoles}, but now taking a cross-section along the diagonal $k_2 = k_3$. We show two equilateral $l$-triplets $l_1 = l_2 = l_3 = l$ with $l = 10$ and 30. The plot is cut at $k_2/k_1 = 0.5$ on the left since there is no signal below it where the triangle condition is violated (this property does not hold for the observed bispectrum in a realistic Universe, however). At high $k_2$, the SFB bispectrum is proportional to the Fourier-space bispectrum.}
\label{fig:iso_homo_1D_equilateral}
\end{figure*}

We also show a one-dimensional cross section in Fig.~\ref{fig:iso_homo_1D_equilateral}, taking the diagonal $k_2 = k_3$, for two different equilateral $l$ shapes ($l_1 = l_2 = l_3=l$). Perhaps the most important feature to note in this plot is that the SFB bispectrum effectively reduces to the matter bispectrum at the limit $k_2=k_3 \gg k_1$ in the homogeneous and isotropic Universe. In this limit, we have $\cos(\theta_{12})=\cos(\theta_{13}) \to 0$ and $B^{\mathrm{iso/homo}}_{l_1l_2l_3}(k_1,k_2,k_3)$ reduces to $C_{l_1l_2l_3}B_m(k_1,k_2,k_3)$ for a prefactor $C_{l_1l_2l_3}$, which quickly tends towards a constant as $l$ grows.

Similarly, in the limit of degenerate triangles e.g. $k_3=k_2+k_1$, we have $\cos(\theta_{12})=\cos(\theta_{13}) =1$, such that only the $m=0$ term in Eq.~\ref{eq:sfb_3point_expanded2} is nonzero, and the bispectrum reduces to 
\ba
B^{\text{SFB, iso/homo}}_{l_1l_2l_3}(k_1,k_2,k_3)
= B_{m}(k_1, k_2, k_3) i^{l_1+l_2+l_3}
\sqrt{2(2l_1+1)}
\begin{pmatrix}
  l_1 & l_2 & l_3 \\
  0 & 0 & 0
\end{pmatrix}
\,,
\label{eq:sfb_iso_homo_degenerate_k}
\ea
where again we have that the cosmological signal and the geometric coupling separate into a $k$-dependent and an $l$-dependent piece.

\subsection{Requirement on the sampling frequency for resolving the oscillations}
\label{sec:freq_oscillations}

As we will see in the next section, the observed SFB bispectrum signal has a similar oscillation pattern in the space of $k$'s as in the isotropic and homogeneous limit. With the analytic formula Eq.~\ref{eq:sfb_3point_expanded2} at hand, we can easily estimate the local frequency of oscillations in $k$-space, and thus the minimum sampling of $k_i$'s required to resolve these oscillations, assuming that the computation is performed on a uniform cubic grid of $(k_1,k_2,k_3)$ with spacing $\Delta k$.

For example, consider for a given $k_1$, the oscillations along the diagonal $k_2=k_3$, as are visible in Fig.~\ref{fig:iso_homo_1D_equilateral}. Estimating the frequency of the oscillations amounts to estimating the spacing between the roots of the associated Legendre polynomials in the products of $Y_{lm}$'s in Eq.~\ref{eq:harmonic_product}. On the diagonal, the lowest point for which the signal is nonzero corresponds to the degenerate isosceles triangle $k_2 = k_3 = k_1/2$, and the diagonal extends to the top right into a squeezed triangle where $k_2, k_3 \gg k_1$. On this trajectory, $\theta_{12}$ and $\theta_{13}$ vary from 0 to $\pi/2$. The associated Legendre polynomial $P^l_m(\cos(\theta))$ has $l-|m|$ roots on the range $0<\theta<\pi$, which are symmetric about $\pi/2$, so there are $(l-|m|)/2$ roots on the ranges we consider.

Consequently, the product of $Y_{l_2 m} Y_{l_3m}$ crosses zero at most $(l_2+l_3-2|m|)/2$ times\footnote{If $l_2=l_3$, it crosses zero $\sim l_2/2$ times.}. The $m = 0$ term has the highest number of roots and may be used to estimate an upper bound on the spacing of roots. Towards that purpose, note that $P^0_l(\cos(\theta))$ is simply the Legendre polynomial $\mathcal{L}_l(\cos(\theta))$. Let $\theta_1, \dots, \theta_l$ be the sequence of roots of $\mathcal{L}_l(\cos(\theta))$ in the interval $(0,\pi)$, listed in increasing order. Then we have the inequalities on the location of the roots~\cite{legendre_roots}
\ba
\frac{\nu-\frac{1}{2}}{l} \pi<\theta_\nu<\frac{\nu}{l+1} \pi \quad(\nu=1,2, \cdots ,\left\lceil l/ 2\right\rceil) \,.
\ea
Hence, the first value of $k_2$ above $k_{2, \rm{min}} = k_1/2$ for which the spherical harmonic product of index $m=0$ vanishes along the diagonal satisfies 
\begin{equation}
k_2^{\rm root\,\nu=1} < \frac{k_1}{2\cos(\pi/(l_3+1))}, 
\end{equation}
Thus, to have at least $N$ sampling points per oscillation, the sampling $\Delta k$ must satisfy
\ba
N \Delta k < k_2^{\rm root\,\nu=1} - k_{2, \rm{min}} < k_1\left[\frac{1}{2\cos(\frac{\pi}{l_3+1})}-\frac{1}{2} \right]\;\xrightarrow{l_3 \gg \pi}\; \frac{k_1}{4}\left(\frac{\pi}{l_3}\right)^2.
\label{eq:resolution_scaling} 
\ea
We have verified numerically that the estimated $l_3^{-2}$ scaling holds for bispectrum signal. Given the cost of computing the observed bispectrum, this means that resolving the oscillations of the signal within the triangle inequality region is challenging at high $l$.

In Section \ref{sec:properties}, we shall see that one property of the observed bispectrum signal is that for large enough $l$ it is ``Limber-suppressed" at low-$k$, in particular inside the region where the triangle inequality is satisfied. This means that we actually do not need to resolve the signal close to the borders of this region, where the local frequency of oscillations is higher, as the signal contains comparatively little information there. 

\section{The observed SFB bispectrum}
\label{sec:observed_sfb_s}

In this section, we begin by describing the template decomposition we use in order to render the computation of the observed SFB bispectrum feasible. We then give details of the signal computation before studying its properties.

Note that we now incorporate redshift evolution, RSD effects, PNG and survey window effects. Statistical homogeneity is now broken by the growth of structure and RSD\footnote{Note that in Fourier space, one can still assume statistical homogeneity by restricting to a given redshift bin and choosing an effective redshift for the entire bin.}, and the latter also breaks intrinsic isotropy. We restrict ourselves to the linear regime $k \lesssim  0.1 \, \iMpch$ for which the tree-level bispectrum remains valid down to $z=0$, and therefore do not include Fingers-of-God effects. We also choose to not model the monopole and dipole here, as they are affected by observer terms in GR such as the observer potential and peculiar velocity \cite{bertacca}, 
i.e we restrict ourselves to multipoles $l_i \ge 2$.

\subsection{Template decomposition of the observed bispectrum}
\label{sec:main_derivation}

We begin by expressing the observed galaxy density field by applying the window function to the galaxy density field in redshift space,
\ba
\delta^{g,\obs}(\vr)
&=
W(\vr)
\int\frac{\dd^3\vq}{(2\pi)^3}\,e^{i\vq\cdot\vr}
\delta^{g}(\vq,\vr) \,.
\label{eq:observed_galaxy_density_general}
\ea
To second order in the
linear matter density field $\delta^{(1)}$, we have from Eq.~\ref{eq:Zn} that $\delta^{g,\obs}(\vr)=\delta^{g,\obs,(1)}(\vr)+\delta^{g,\obs,(2)}(\vr)$, where 
\ba
\delta^{g, \obs,(1)}(\vr)
&=
W(\vr)
\int\frac{\dd^3\vq}{(2\pi)^3}\,e^{i\vq\cdot\vr}
\,D(r)\,Z_1(\bm{q},\vr)\,\delta^{(1)}(\bm{q})\,,
\label{eq:deltag1}
\\
\delta^{g, \obs,(2)}(\vr)
&=
W(\vr)
\int\frac{\dd^3\vq}{(2\pi)^3}\,e^{i\vq\cdot\vr}
\,D^2(r)\int \frac{1}{(2\pi)^3} \dd^3\vk_1\,\dd^3\vk_2\,Z_2(\vk_1,\vk_2,\vr)\,\delta^{(1)}(\vk_1)\delta^{(1)}(\vk_2)\,\delta_D(\vk_1+\vk_2-\vq)\,.
\label{eq:observed_galaxy_density_order2}
\ea

Transforming the linear density contrast Eq.~\eqref{eq:deltag1} into SFB space, we retrieve Eq.~\eqref{eq:sfb_coeff_obs}
where the kernel $\mathcal{W}_{\ell m}^{LM}(k,q)$ encodes (linear) galaxy physics and RSD. We aim to derive a similar relation for the second-order density contrast.
We now transform Eq.~\ref{eq:observed_galaxy_density_order2} into SFB space using Eq.~\ref{eq:sfb_fourier_pair_b}. Expressing the linear matter density contrast in the SFB basis from Eq.~\ref{eq:delta_from_deltalm}, and writing the Dirac-delta as an integral over complex exponentials, we obtain
\ba
\delta_{\ell m}^{g,(2)}(k)
&=
\sqrt{\frac{2}{\pi}}\,k
\int\dd^3\vr
\,j_\ell(kr)\,Y^*_{\ell m}(\rhat)
\,W(\vr) \,D^2(r)
\int\dd^3 \vq\,e^{i\vq\cdot\vr}
\int\frac{\dd^3\vk_1}{(2\pi)^3}\int\frac{\dd^3\vk_2}{(2\pi)^3}
\,Z_2(\vk_1,\vk_2,\vr)
\int\dd^3\vx
\,e^{i\vk_1\cdot\vx}
\,e^{i\vk_2\cdot\vx}
\,e^{-i\vq\cdot\vx}
\vs&\quad\times
\frac{1}{k_1}
\sum_{L_1M_1}
i^{-L_1}
\,Y_{L_1 M_1}(\khat_1)
\,\delta^{(1)}_{L_1 M_1}(k_1)
\frac{1}{k_2}
\sum_{L_2M_2}
i^{-L_2}
\,Y_{L_2 M_2}(\khat_2)
\,\delta^{(1)}_{L_2 M_2}(k_2)\,.
\label{eq:delta_2_expanded}
\ea

Naively inserting the expression for $Z_2$ (Eq.~\ref{eq:z2_with_fnl}) into Eq.~\ref{eq:delta_2_expanded} would require evaluating high-dimensional angular integrals, which is intractable. To simplify the calculation, we remark that $Z_2(\vk_1,\vk_2,\vr)$ is nearly
a polynomial in $\khat_1\cdot\khat_2$, $\khat_1\cdot\rhat$, and $\khat_2\cdot\rhat$. Indeed, defining $\tilde Z_2$ such that
\ba
Z_{2} (\vk_1,\vk_2,\vr)&=\tilde{Z}_2(\vk_1,\vk_2,\vr) + f_{\mathrm{NL}} \frac{\alpha(k)}{\alpha\left(k_{1}\right) \alpha\left(k_{2}\right)}\left(b^\mathrm{E}_{10} + f(r)\mu^2\right) +f(r) \mu^{2}G_{2}\left(\vk_{1}, \vk_{2}\right)
\label{eq:z2_with_fnl_split} \,,
\ea
we can decompose $\tilde{Z}_{2}$ into Legendre polynomials in those three variables and thereby factorize the dependence on the $\khat_i$ and $\rhat$:
\ba
\tilde{Z}_2(\vk_1,\vk_2,\vr)
&=
\sum_{l_1l_2l_3} Z_{l_1 l_2 l_3}(k_1,k_2,r)
\,\mathcal{L}_{l_1}(\khat_1\cdot\khat_2)
\,\mathcal{L}_{l_2}(\khat_1\cdot\rhat)
\,\mathcal{L}_{l_3}(\khat_2\cdot\rhat)
\,.
\label{eq:z2_decomposed}
\ea
Importantly, the sum over $l_1,l_2,l_3$ is  finite, and indexed by  $9$ triplets $(l_1,l_2,l_3)$ whose corresponding coefficients $Z_{l_1l_2l_3}$ are listed in \cref{app:Zlll_terms}. As we  discuss below, the above Legendre decomposition permits to reduce the bispectrum to a triple integral. 

Two other terms remain. The term proportional to $f_{\mathrm{NL}}$  in Eq.~\ref{eq:z2_with_fnl_split} depends on $k=|\vk_1+\vk_2|$ through $\alpha(k)$, so it cannot be decomposed it in a similar fashion. In principle, the term proportional to $G_{2}(\vk_{1}, \vk_{2})$ in Eq.~\ref{eq:z2_with_fnl_split} can be decomposed in this manner; however, since $\mu^2=(\vk_1\cdot\rhat + \vk_2\cdot\rhat)/(k_1^2+k_2^2+2\vk_1\cdot\vk_2)$ it would render the sum Eq.~\ref{eq:z2_decomposed} infinite and slowly-converging. Hence, we choose to treat the $G_2$ and $f_{\mathrm{NL}}$ terms separately (and exactly), as described in \cref{app:velocity_div_kernel}.

We summarize the remainder of the derivation here, leaving details to Appendix \ref{app:bispectrum_comutation_details}: First, we insert the decomposition Eq.~\ref{eq:z2_decomposed}  of $\tilde{Z}_2(\vk_1,\vk_2,\vr)$ into Eq.~\ref{eq:delta_2_expanded}, and use the plane-wave expansion Eq.~\ref{eq:rayleigh} to decompose the complex exponentials into spherical harmonics and spherical Bessel functions. We rid of the angular integrals over $\qhat$ with the orthogonality relation for spherical harmonics Eq.~\ref{eq:YlmYlmDelta}. 

Then we apply Wick's theorem to compute 
$\<
\delta_{\ell m}^{g,(2)}(k)
\,\delta^{g,(1)}_{\ell' m'}(k')
\,\delta^{g,(1)}_{\ell'' m''}(k'')
\>$ in terms of the two-point functions. Finally, proceeding under the assumption of a spherically symmetric window $W(\vr)=W(r)$, we compute the angle averaged bispectrum using Eq.~\ref{eq:sfb_3point_reduced} and obtain:
\ba
B^{\text{SFB}}_{l_1l_2l_3}(k_1,k_2,k_3)&=
2\int\dd q_2
\mathcal{W}_{l_2}(k_2,q_2)
\,P(q_2)
\int\dd q_3
\,\mathcal{W}_{l_3}(k_3,q_3)
\,P(q_3) \ \mathcal{V}_{\mathrm{tot.}}^{l_1l_2l_3}(k_1,q_2,q_3)+ \mathrm{2\ cyc.\  perm.} \,,
\label{eq:sfb_tot}
\ea
where
\ba
\mathcal{V}_{\mathrm{tot.}}^{l_1l_2l_3}(k_1,q_2,q_3)&\equiv
\mathcal{V}_{\mathrm{}}^{l_1l_2l_3}(k_1,q_2,q_3)+
\mathcal{V}_{f_{\mathrm{NL},G_2}}^{l_1l_2l_3}(k_1,q_2,q_3) \,,
\label{eq:wllltot} 
\ea
where the specific forms of $\mathcal{V}_{\mathrm{}}^{l_1l_2l_3}$ and $\mathcal{V}_{f_{\mathrm{NL},G_2}}^{l_1l_2l_3}$ are given by Eqs.~\ref{eq:wlll_averaged} and~\ref{eq:wlllG2} respectively.
Note that the kernel $\mathcal{W}_l(k,q)$ is already given by Eq.~\ref{eq:W_lkq_first}.

Let us briefly comment on the form of the dimensionless kernel Eq.~\ref{eq:wllltot}. The first term  of Eq.~\ref{eq:wllltot} is given by 
\ba
\mathcal{V}^{\ell L_1 L_2}(k,k_1,k_2) \equiv (32\pi)^{\frac{3}{2}}kk_1k_2
\sum_{l_1l_2l_3L_3L_4}g^{L_1 L_2 \ell}_{l_1l_2l_3L_3L_4} J^{\ell L_3L_4}_{l_1l_2l_3}(k,k_1,k_2) \,,
\label{eq:wlll_averaged_main_text}
\ea
where 
\begin{align}
J^{\ell L_3L_4}_{l_1l_2l_3}(k,k_1,k_2)\equiv
\int\dd r\,r^2
\,j_\ell(kr)
\,j_{L_3}(k_1r)
\,j_{L_4}(k_2r)
\,W(r) \,D^2(r)
\,Z_{l_1 l_2 l_3}(k_1,k_2,r) \,,
\label{eq:J_integral_main_text}
\end{align}
is the contribution to SFB mode coupling by cosmological sources (e.g. redshift evolution, RSD and PNG) and survey window, and where $g^{L_1 L_2 \ell}_{l_1l_2l_3L_3L_4}$ is a purely geometric mode coupling coefficient given by Eq.~\ref{eq:gcoeffdef}. 

In analogy to the SFB power spectrum, in which the matter power spectrum is convolved with kernels $\mathcal{W}_{\ell}(k,q)$ that describe the mode coupling through the product of two spherical Bessel functions, the kernel $\mathcal{V}^{l_1l_2l_3}(k_1,q_2,q_3)$ contains a product of three spherical Bessel functions. This endows the SFB bispectrum with key geometric features which we discuss shortly. The second term of Eq.~\ref{eq:wllltot}, given by Eq.~\ref{eq:sfb_G2_appendix}, contains the contribution from the $f_{\mathrm{NL}}$ and $G_2$ terms and is of a similar form to Eq.~\ref{eq:wlll_averaged_main_text}.  Our result matches the general form of the SFB bispectrum derived in Ref.~\cite{bertacca}. 
\subsection{Signal computation}
\label{sec:observed_sig_modeling_results}

We compute the bispectrum for a uniform grid of $(k_1,k_2,k_3)$ of size $200^3$ with $k_i$ between $k_\mathrm{min}=4 \times 10^{-3}\ \iMpch$ (note that future surveys like SPHEREx will be able to probe down to $10^{-3}\ \iMpch$) and $k_\mathrm{max}= 8\times 10^{-2} \ \iMpch$ (to stay within linear regime) with uniform spacing $ \Delta k =3.8 \times 10^{-4}  \ \iMpch$. For the toy window function, we assume a sphere $W(\vr)=\mathbf{1}_{[0,r_\mathrm{max}]}(r)$ with $r_\mathrm{max}=5000 \,\Mpch$, corresponding to a maximum redshift $z \sim 4.1$. As a result of the large redshift range chosen here, the kernels $\mathcal{W}_l(k,q)$ in Eq.~\ref{eq:sfb_tot} are highly peaked around $k \approx q$. For surveys with a smaller redshift extent $r_\mathrm{max}$, the kernel $\mathcal{W}_l(k,q)$ would have a smoother peak and lower frequency oscillations (as in Fig.~\ref{fig:Wlkq}), which would make the computation less computationally demanding. 

Let us now examine more closely the form of the integrals to be computed. Note first that Eq.~\ref{eq:sfb_tot} is a two-dimensional integral (over $q_2$ and $q_3$) of the kernels $\mathcal{V}_{\rm tot}$. 
The first term in this kernel, $\mathcal{V}^{l_1 l_2 l_3}$ (Eq.~\ref{eq:Vlll_def}) is a sum of the one-dimensional Bessel integrals given in Eq.~\ref{eq:J_integral_main_text}. The second term $\mathcal{V}^{l_1 l_2 l_3}_{\fnl, G_2}$ is also effectively a one-dimensional integral since the computationally-intensive parts of the integrand, the integrals $\mathcal{W}^{G_2}_l$, $\mathcal{W}^{f_\mathrm{NL}}_l$ and $I_{l_1l_2l_3}$, can be precomputed on a grid. 
The precomputation for $I_{l_1l_2l_3}$ requires a few seconds using the identity Eq.~\ref{eq:gauntlike_identity}, which expresses it as a finite sum with $l_3$ terms. 

Finally, the kernels $\mathcal{V}^{l_1 l_2 l_3}(k_1, q_2, q_3)$ themselves are precomputed on a $q_2$-$q_3$ grid to be reused for the various triplets $(k_1,k_2,k_3)$. This, along with the final integration in Eq.~\ref{eq:sfb_tot}, is the bottleneck of the calculation, and limits the number of $l$-triplets we may feasibly calculate. However, the signal is sufficiently smooth in $l$ that this might not pose a problem for e.g, a Fisher forecast exercise. All grid computations are parallelized using Julia's multithreading functionality; computing the kernels $\mathcal{V}^{l_1 l_2 l_3}(k_1, q_2, q_3)$ requires a few hours per triplet $(l_1,l_2,l_3)$ with multithreading across $256$ AMD EPYC 7763 CPUs.

Beyond dimensionality, a second numerical concern is the oscillatory nature of the integrands. We perform all integrals via Gauss-Legendre quadrature. To accurately integrate $\mathcal{V}_\mathrm{tot}^{l_1l_2l_3}(k_1,q_2,q_3)$, we observe (empirically) that if the averaging $r$-spacing is $\Delta r$, then we must impose $k_1+q_2+q_3 \lesssim \pi/\Delta r$. 
We use $\Delta r=5 \ \Mpch$. 
Further, to evaluate the bispectrum we must convolve the kernels $\mathcal{V}_\mathrm{tot}^{l_1l_2l_3}(k_1,q_2,q_3)$ with the $\mathcal{W}_{l_2}(k_2,q_2)$, both of which oscillate quickly as $q_2$ 
is varied.
This requires a sufficient sampling in the $q_i$ space. 
As the computation time is quadratic in the number of sampling points for each $q_i$, only modest samplings are feasible; we use $300$ points for each $q_i$, chosen as Gauss-Legendre nodes. 

\subsection{Properties of the signal}
\label{sec:properties}

The observed bispectrum signal displays a number of salient features which we now discuss. 

\subsubsection{Oscillations in $k$ and mode coupling}

\begin{figure*}[!t]
\centering
\includegraphics[width=0.98\linewidth]{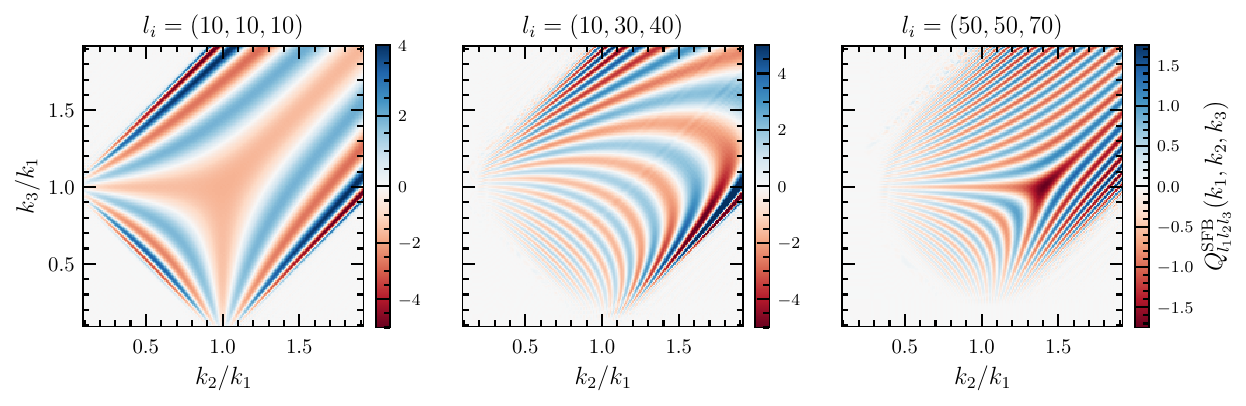}
\caption{The observed reduced SFB bispectrum for a realistic Universe, assuming a spherically symmetric survey window, for the same set of $l$-triplets as in Fig.~\ref{fig:iso_homo_mixed_multipoles}. 
The signal is sampled at $200^2$ pairs $(k_2,k_3)$ in each panel, whereas in Fig.~\ref{fig:iso_homo_mixed_multipoles} there are $400^2$ sampled pairs.
}
\label{fig:observed_3_panel}
\end{figure*}

A cross-section of the reduced bispectrum for fixed $k_1=0.0418 \, \iMpch$, for the same set of $l$-triplets as in Fig.~\ref{fig:iso_homo_mixed_multipoles}, is shown in Fig.~\ref{fig:observed_3_panel}. Perhaps the most striking feature here is that the patterns of oscillations in $k$-space are similar to those visible in Fig.~\ref{fig:iso_homo_mixed_multipoles} for the isotropic and homogeneous case. We may understand this from the mode coupling coefficients $g^{L_1 L_2 \ell}_{l_1l_2l_3L_3L_4}$ in Eq.~\ref{eq:wlll_averaged_main_text}. Numerically we find that they are generally suppressed unless $(L_3 ,L_4)=(L_1 ,L_2)$, such that the dominant contribution to the bispectrum signal $B^\mathrm{SFB}_{l_1l_2l_3}$ is from integrals of type Eq.~\ref{eq:J_integral_main_text} where the spherical Bessel functions have the same indices  $l_1,l_2,l_3$, as in the isotropic and homogeneous case (Eq.~\ref{eq:sfb_3point_expanded1} --~\ref{eq:gauntlike_propto_gaunt_main_text}).

\subsubsection{Limber suppression at low $k$ and high $l$}

\begin{figure*}[ht]
\centering
\includegraphics[width=0.6\linewidth]{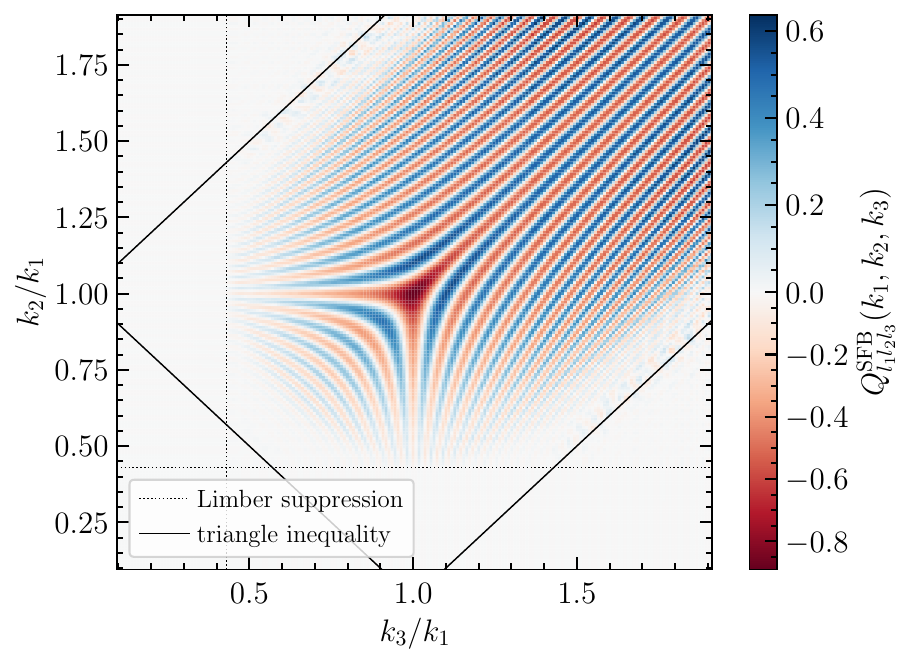}
\caption{A illustration of the Limber suppression at low $k$, which becomes visible for the reduced observed SFB bispectrum signal within the triangle inequality region for sufficiently large $l$. Here we show the signal for $l_1=l_2=l_3=90$,  while still fixing $k_1=4.18 \times 10^{-2} \, \iMpch$. The onset of Limber suppression is indicated by the gray dashed lines, where we expect the signal to be suppressed according to the Limber approximation $k_i=(l_i+\frac{1}{2})/r_\mathrm{max}$. The border of the region where the triangle inequality on $(k_1,k_2,k_3)$ holds is shown by the solid lines. 
}
\label{fig:equilateral_2D_ell=90}
\end{figure*}

In \cref{fig:equilateral_2D_ell=90}, we increase the $l$ values to $l_1 = l_2 = l_3 = 90$, and see that the overall amplitude of the oscillations decreases by roughly an order of magnitude relative to the leftmost panel of Fig.~\ref{fig:observed_3_panel}. To understand the suppression with increasing $l$, which is generic, we note that for fixed $r$, the spherical Bessel function $j_l(kr)$ is proportional to $(kr)^l$ for small $k$; for large $k$, it oscillates with an amplitude proportional to $(kr)^{-1}$. Further, in the Limber approximation\footnote{Calling this the \emph{Limber approximation} is standard in cosmology. However, the term is slightly misleading, since the original approximation by \citet{1954ApJ...119..655L} was in configuration space, and only applied to Fourier-space by \citet{1998ApJ...498...26K}. The resulting approximation effectively is that of Eq.~\ref{eq:jl_limber} \citep{2008PhRvD..78l3506L}.}
(Eq.~\ref{eq:jl_limber}), for a fixed $l$ and $k$, the Bessel function $j_l(kr)$ is peaked around $r \sim (l+\frac{1}{2})/k$ 
with a peak value equal to $\sim \sqrt{\pi/(2l)}$ \citep{2008PhRvD..78l3506L}.

Physically, we may understand the suppression as follows. For fixed $(k_1,k_2,k_3)$, the bispectrum at higher $l_i$ probes higher redshifts. If the survey window has finite radial extent, these higher redshift contributions to the signal are necessarily smaller, and vanish once the redshift exceeds the extent of the survey. By contrast, when the survey window has infinite size as in the homogeneous and isotropic Universe, for every $l_i$ there is a corresponding redshift which contributes non-negligibly to the signal, such that there is no such suppression at high $l_i$.

Given that the spherical Bessel functions go as $(kr)^{l}$ when $k \lesssim l/r$, 
we should also expect a sharp suppression of the bispectrum at low $k$. 
This suppression is not visible in the first panel of  Fig.~\ref{fig:observed_3_panel}
because this effect is only relevant for $l \gtrsim k_\mathrm{min}r_\mathrm{max}=20$ in our fiducial setup.  
On the other hand it visible in Fig.~\ref{fig:equilateral_2D_ell=90}, for $l_1 = l_2 = l_3 = 90$. The onset of the low $k$ suppression is inside of the region where the triangle inequality holds, and is indicated by the dotted gray lines. 

Recall from the previous sections that the frequency of the oscillations increases as we approach the triangle inequality boundary, making the computation increasingly difficult close to the boundary with higher sampling needed to resolve these oscillations. For small $l$, we have large spacings $\Delta k \propto l^{-2}$ which are manageable. For large $l$, the Limber suppression is helpful in the sense that it is not necessary to compute the signal at the boundary since it is suppressed there by several orders of magnitude. This is true both in a Fisher analysis or in a real data analysis where we can simply choose to ignore this part of the data vector as it contains almost no information.

\subsubsection{Violation of the triangle condition}
\label{sec:triangle_condition_violation}

We saw that in an isotropic and homogeneous Universe, the bispectrum signal vanishes identically when $(k_1,k_2,k_3)$ does not satisfy the triangle inequality. This is a consequence of (intrinsic) isotropy. In the case of the observed bispectrum, where this isotropy is broken by e.g, redshift space distortions, this is no longer true, though it holds approximately for wider survey windows.  For lower values of $r_\mathrm{max}$, the signal strength is non-negligible even when $(k_1,k_2,k_3)$ violate the triangle inequality.

\begin{figure*}[!t]
\centering
\includegraphics[width=0.98\linewidth]{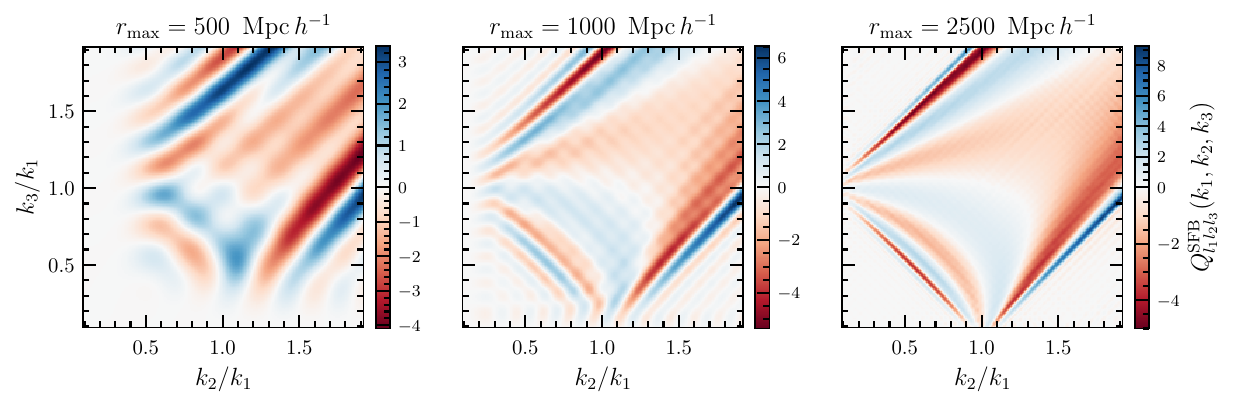} 
\caption{A demonstration of the triangle condition violation in a realistic Universe -- the observed reduced bispectrum signal for different values of the redshift extent $r_\mathrm{max} = 500, 1000$ and $2500\,\Mpch$ (corresponding to $z \sim 0.18$, $0.37$ and $1.2$ respectively) at fixed $k_1=4.18 \times 10^{-2} \ \iMpch$ and $(l_1,l_2,l_3) = (4,6,8)$. For smaller $r_{\rm max}$, the violation of the triangle condition is more severe due to less peaked $\mathcal{W}_l$ kernels (see Section~\ref{sec:triangle_condition_violation} for more explanations).
}
\label{fig:rmax_3_panel}
\end{figure*}

This is due to two reasons. First,  the kernels $\mathcal{W}_l(k,q)$ (Eq.~\ref{eq:W_lkq_first}) are less peaked for smaller $r_\mathrm{max}$, such that for $(k_1,k_2,k_3)$ which violate the triangle inequality, the integral Eq.~\ref{eq:sfb_tot} can pick up a non-negligible contribution from the kernel $\mathcal{V}^{l_1l_2l_3}_{\rm tot}(k_1,q_2,q_3)$ for $(k_1,q_2,q_3)$ which do form a triangle. Secondly, for smaller $r_\mathrm{max}$, $\mathcal{V}^{l_1l_2l_3}_{\rm tot}(k_1,q_2,q_3)$ itself can take non-negligible values for $(k_1,q_2,q_3)$ which do not form a triangle (indeed, the Bessel integrals Eq.~\ref{eq:J_integral_main_text} do not come with any triangle condition). As a result, the triangle inequality is broken in the observed SFB bispectrum, with the violation more severe at lower $r_{\rm max}$. See Fig.~\ref{fig:rmax_3_panel} which illustrates this effect using $r_{\rm max} = 500, 1000$ and $2500$ $\,\Mpch$.

\section{Discussion}
\label{sec:conclusion} 

In this paper, we computed the SFB bispectrum signal for the first time, discussing how to account for redshift space distortions and primordial non-Gaussianity. Starting with a toy example of the homogeneous and isotropic Universe, we built up intuition for later understanding some key features of the observed bispectrum convolved with a toy spherically symmetric window.

To render the computation tractable, we leveraged a decomposition of the second-order redshift space kernel into products of three Legendre polynomials, which allowed us to express the bispectrum, modulo RSD and PNG terms, as a triple integral. Furthermore, we derived an identity to express as a simple sum the $6$-dimensional angular integral of three spherical harmonics (or equivalently, the one-dimensional integral of a product of three spherical Bessel functions on an infinite interval). This enabled us to rapidly compute and study the signal in the case of a homogeneous and isotropic Universe, and to accelerate the calculation of the RSD and PNG contributions to the observed bispectrum.

Even with these techniques and the various numerical optimizations we employed, computing the SFB bispectrum is clearly expensive: for each triplet of multipole indices  $(l_1,l_2,l_3)$ and of wavenumbers $(k_1,k_2,k_3)$, we need to evaluate triple integrals with oscillatory integrands. Our method requires $\mathcal{O}(100)$ CPU hours for each $l$-triplet to compute the signal on a grid of $200^3$ $(k_1,k_2,k_3)$ triplets. In a realistic data analysis, one would need to calculate the signal for different cosmologies in a Monte-Carlo Markov Chain (MCMC) on the order of seconds. We note however that we have chosen a very large redshift extent $z \lesssim 4$, corresponding to $r_{\rm max} = 5000\, \Mpch$, for which the integration is the most challenging. Surveys with smaller redshift extent would require less time for computation. 

There are several possibilities to accelerate and improve the accuracy of the computation that we leave for future work. For example, as the local frequency of the oscillations in the signal can be estimated as a function of $(k_i,l_i)$ from the isotropic and homogeneous bispectrum as in Section \ref{sec:freq_oscillations}, the signal can instead be sampled on a suitable non-uniform grid of $(k_1,k_2,k_3)$. To improve upon the Gauss-Legendre quadrature method for integrating the spherical Bessel product Eq.~\ref{eq:J_integral_main_text}, the $3$-dimensional generalization of the FFTLog method of \cite{Grasshorn_Gebhardt_2018} could prove superior. Leveraging cache-friendly memory layouts may also help to speed up some of the linear algebra operations involved in computing these Bessel integrals on a grid in $k$-space. 

For the purpose of calculations in a MCMC analysis, one could also explore decomposing the bispectrum dependence on cosmological parameters into precomputed templates, and varying the coefficients of the template corresponding to the varying cosmology. If it becomes impossible to directly compute the signal for each point in the cosmology parameter space sampled during the MCMC, the use of emulators could also aid in minimizing the evaluation time. 

Another natural extension to this work is to incorporate more physical and observational effects in the computation. The most important physical effects on large-scales that we have not included here are general-relativistic (GR) effects. The authors of Ref.~\cite{bertacca} detailed how to incorporate them in the SFB bispectrum; in principle they may be evaluated numerically within the same framework described in our work, i.e through (many) additional terms in the first order kernels $\mathcal{W}_l(k,q)$ and in the second-order kernels $\mathcal{V}^{l_1l_2l_3}$, after a template decomposition into Legendre polynomials as in Section \ref{sec:main_derivation}.
On smaller scales, more detailed modeling of RSD (e.g. Fingers-of-God and Alcock-Paczynski effects) and of the nonlinear regime would be needed. In addition, we have used a spherically symmetric window function to demonstrate the calculation, while realistic window convolution is still to be explored.

To feasibly use the SFB bispectrum to analyze survey data, a number of missing pieces would still need to be filled in. In particular, it would be necessary to develop an efficient SFB bispectrum estimator, e.g, by building off of techniques developed by \cite{gebhardt2021superfab, Gebhardt:2023kfu} for an SFB power spectrum estimator.
As allowing for a survey window of arbitrary geometry in the modeling of the signal would greatly increase the computational cost, one may explore accounting for it in the estimator, e.g, by using a windowless estimator which directly returns the window-deconvolved bispectrum as pioneered in Ref.~\cite{Philcox_2021} for the bispectrum multipoles. 

Moreover, a realistic covariance matrix for the SFB bispectrum beyond the Gaussian approximation also needs to be developed, including complexities due to window function convolution as well as non-Gaussian covariance. If the window effects can be reliably removed at the estimator level, then the covariance would be significantly simplified. For the non-Gaussian covariance, an approximation similar in form to that proposed in Ref.~\cite{Biagetti_2022} may be applicable for the SFB bispectrum, where the non-Gaussian part of the covariance is dominated by the product of two bispectra sharing the same large scale -- a good approximation for squeezed configurations and also tested to be good enough for other configurations in Ref.~\cite{Biagetti_2022} in the context of Fourier-space bispectrum.  Alternatively, to incorporate all complexities at once, one may also develop mocks to compute the mock-based covariance by averaging over many realizations, once a fast SFB bispectrum estimator exists. This method would include wide-angle effects directly for mocks with large enough angular area, while it could be challenging to incorporate all GR effects into the mocks.

Since an advantage of the SFB formalism is that it avoids the loss of information due to assuming inexact lines-of-sights for individual galaxy triplets, it would also be interesting to evaluate more quantitatively now this information gain, for example by comparing to the standard bispectrum multipole formalism in the local plane-parallel approximation and to perturbative corrections thereof as in \cite{pardede2023wideangle}. Our work to enable the computation of the signal will allow for such a study to be conducted. With a suitable scheme to interpolate the signal in the space of multipoles $l_i$, it could be feasible to conduct a Fisher forecast for various cosmological parameters of interest, such as $\fnl$ or RSD parameters.

Note that this loss of information may be small for surveys with small angular extent, but more important for full-sky surveys like SPHEREx. Currently, with the exception of the TSH formalism, only perturbative approaches to modeling wide-angle effects in the bispectrum have been proposed \cite{pardede2023wideangle, Noorikuhani:2022bwc}, expanding from the global plane-parallel approximation. Thus, the SFB bispectrum remains the only method to fully account for all large scale effects non-perturbatively on the largest angular scales while preserving the potential of retaining all information contained in the radial modes.

Another advantage of the SFB formalism is that some of the GR terms, which are mostly radial effects along the line-of-sight, become easily disentangled from other effects. In particular, the monopole and the dipole terms in the SFB formalism contain all the observer terms in GR arising from the potential and velocity at the observer position. Some of these terms may be quantified via other means before being subtracted (e.g. the velocity term in the dipole), while others are intrinsically undetectable (e.g. observer potential) and may need to be modeled through constrained realization if they affect observables of interest. 

Modeling these terms in a Cartesian framework amounts to propagating these effects to every mode (and every order if a perturbative expansion from the plane-parallel approximation is used), which would propagate potential systematics into every measured mode. In contrast, in a spherical framework such as SFB, there is a clear radial and angular separation that allows for the isolation of such terms into just the monopole and dipole, which may then be discarded or tested separately for systematics. 

While the TSH formalism provides a similar advantage, it requires many radial bins required to resolve the large scale radial modes (which is important to do for measuring $\fnl$), which introduces highly correlated neighboring radial bins, and leads to numerically instabilities during covariance inversion. The SFB method is therefore a trade-off between extracting the maximal amount of information and the cost of computing the signal. In this regard, we have made a step forward by rendering the SFB signal computable and studying its various features. 

This is merely the beginning of more efforts to follow to make the calculation of the SFB bispectrum feasible for next-generation surveys. With increasing computational power in the future, along with more sophisticated numerical and mathematical techniques, the SFB bispectrum may become a key formalism that will allow us to extract all of the possible information from a full-sky galaxy survey.

\section{ Acknowledgements}
This research used resources of the National Energy Research Scientific Computing Center (NERSC). We thank Johannes Blaschke for invaluable technical advice on running and optimizing our Julia code on NERSC,  Vincent Desjacques for enlightening discussions, and Robin Wen for helpful feedback on the manuscript. J.B. is supported in part by the DOE Early Career Grant DESC0019225. A.T. is also grateful for the support provided by the Walter Burke Institute for Theoretical Physics. C.H. thanks her Maker for providing the amazing people to work together, the opportunities to collaborate fruitfully, and all the resources and any insights that made this work possible. This material is based upon work supported by the U.S. Department of Energy, Office of Science, Office of High Energy Physics, under Award Number DE-SC0011632. 

\bibliography{sfb.bib}

\appendix
\clearpage
\section{Useful formulae}
\label{app:sfb_useful_formulae}

\subsection{Dirac delta distribution}
For a continuously differentiable function $g$ with simple roots $\{x_i\}$ we have 
\ba 
\label{eq:dirac_composition}
\delta_D(g(x))=\sum_i\frac{\delta_D(x-x_i)}{|g'(x_i)|} \,.
\ea
For any function $f(\vk)$,
\ba
\int k^2\dd{k}\,\dd^2\khat\,\delta^D(\vk-\vk')\,f(\vk)= \int\dd{k}\,\delta^D(k-k')\int\dd^2\khat\,\delta^D(\khat-\khat')\,f(\vk)\,.
\ea
Therefore,
\ba
\label{eq:dirac3D}
\delta^D(\vk-\vk')
&= k^{-2} \, \delta^D(k-k')\,\delta^D(\khat-\khat')\,.
\ea
\subsection{Spherical Bessel functions}
To first order the  Bessel function $J_\nu(x)$ may be approximated by a Dirac delta as \cite{2008PhRvD..78l3506L} 
\ba
J_\nu(kr) &\simeq \delta^D\!\(kr-\nu\)\,.
\ea
Therefore, for a spherical Bessel function
$j_\ell(x)=\sqrt{\pi/2x}\,J_{\ell+\frac12}(x)$ we have
\ba
\label{eq:jl_limber}
j_\ell(kr)
&\simeq \sqrt{\frac{\pi}{2rk}}\,\delta^D\!\(r-\frac{\ell+\frac12}{k}\) \,,
\ea
to first order.  In the cosmology literature, a version of this is often called \emph{Limber's approximation}
\citep{1954ApJ...119..655L}.
Spherical Bessel functions satisfy the orthogonality relation
\ba
\label{eq:jljlDelta}
\delta^D(k-k')
&= \frac{2kk'}{\pi}\int_0^\infty\dd{r}\,r^2\,j_\ell(kr)\,j_\ell(k'r)\,.
\ea
\subsection{Spherical harmonics}
Spherical harmonics can be expressed in terms of a complex exponential and real
associated Legendre functions $\mathrm{P}_\ell^m(x)$ as
\ba
\label{eq:spherical_harmonics}
Y_{\ell m}(\rhat)
&=
e^{im\phi}
\(\frac{(\ell - m)! (2\ell+1)}{4\pi\(\ell+m\)!}\)^\frac12
\mathrm{P}_\ell^m\!\(\cos\theta\).
\ea
The associated Legendre functions are even or odd according to the index,
\ba
P_{\ell}^m(-x)=(-1)^{\ell+m} P_{\ell}^m(x) \,.
\label{eq:plm_parity}
\ea
The completeness relation is
\ba
\label{eq:Ylm_completeness}
\sum_{\ell m} Y_{\ell m}(\rhat)\,Y^*_{\ell m}(\rhat')
&= \delta^D(\rhat - \rhat')\,.
\ea
The spherical harmonics satisfy the orthogonality relation
\ba
\label{eq:YlmYlmDelta}
\int\dd{\Omega}_{\rhat}\,Y_{\ell m}(\rhat)\,Y^*_{\ell'm'}(\rhat)=\delta^K_{\ell\ell'}\delta^K_{mm'} \,.
\ea
For a rotation $\mathcal{R}$ about the origin that sends the unit vector $\mathbf{r}$ to $\mathbf{ r}'$, we have
\ba
Y_{\ell,m}(\mathbf{r}') = \sum_{m' = -\ell}^\ell [D^{(\ell)}_{mm'}({\mathcal{R}})]^* Y_{\ell,m'}({\mathbf{r}}),
\label{eq:rotation_ylm}
\ea
where $[D^{(\ell)}_{mm'}({\mathcal{R}})]^*$ is the complex conjugate of an entry of the Wigner $D$-matrix.
The Wigner $D$-matrix is a unitary square matrix of dimension $2j+ 1$. If $\mathcal{R}$ is defined by proper Euler angles $\alpha,\beta,\gamma$ in the $z$-$y$-$z$ convention, we have the property 
\ba
D^{\ell}_{m 0}(\mathcal{R}) = \sqrt{\frac{4\pi}{2\ell+1}} Y_{\ell,m}^{*} (\beta, \alpha ) \,,
\label{eq:WignerD_lm0}
\ea
and also the relation
\ba
  D^j_{m k}(\mathcal{R}) D^{j'}_{m' k'}(\mathcal{R}) =
  \sum_{J=|j-j'|}^{j+j'}  \langle j m j' m' | J \left(m + m'\right) \rangle
               \langle j k j' k' | J \left(k + k'\right) \rangle
  D^J_{\left(m + m'\right) \left(k + k'\right)}(\mathcal{R})
  \label{eq:WignerD_pairwise_prod} \,,
\ea
where  $\langle j_1 m_1 j_2 m_2 | j_3 m_3 \rangle$ is a Clebsch-Gordan coefficient. The latter is related to the Wigner $3j$ symbols by:
\ba
 \langle j_1 \, m_1 \, j_2 \, m_2 | J \, M \rangle
  &= (-1)^{-j_1 + j_2 - M} \sqrt{2 J + 1}
    \begin{pmatrix}
      j_1 & j_2 &  J \\
      m_1 & m_2 & -M
    \end{pmatrix} .
\label{eq:WignerD_toWigner3j}
\ea
We may also expand plane waves in terms of spherical Bessels and
spherical harmonics,
\ba
\label{eq:rayleigh}
e^{i\vq\cdot\vr} &=
4\pi\sum_{\ell',m'} i^{\ell'} j_{\ell'}(qr)\,
Y^*_{\ell'm'}(\qhat)\,Y_{\ell'm'}(\rhat)\,,
\ea
from which it follows, using Eq.~\ref{eq:YlmYlmDelta}, that
\ba
\int \dd^2\rhat Y^*_{l,m}(\rhat)e^{i\vq \cdot\vr}(\qhat \cdot \rhat)^{\alpha}=4\pi i^l Y^*_{l,m}(\qhat)(-i\partial_{qr})^{\alpha}j_l(qr) \,.
\label{eq:spherical_transform_g1}
\ea
The Legendre polynomials can be expressed as a sum over spherical harmonics as
\ba
\label{eq:legendre_spherical_harmonics}
\mathcal{L}_\ell(\khat\cdot\rhat)
&= \frac{4\pi}{2\ell+1}\sum_m Y_{\ell m}(\khat)\,Y_{\ell m}^*(\rhat)\,.
\ea
We also have the identities
\ba
\label{eq:Ylm_conjugate}
Y^*_{\ell m}(\rhat)
&= (-1)^m Y_{\ell,-m}(\rhat)
\,,\\
\label{eq:Ylm_parity}
Y_{\ell m}(-\rhat)
&= (-1)^\ell Y_{\ell,m}(\rhat)\,.
\ea
\subsection{Gaunt factor}
The Gaunt factor is
\ba
\label{eq:gaunt_factor}
\mathcal{G}^{\ell L L_1}_{mMM_1}
&\equiv
\int\dd^2\rhat
\,Y_{\ell m}(\rhat)
\,Y_{LM}(\rhat)
\,Y_{L_1M_1}(\rhat)\,,
\ea
and it can be expressed in terms of Wigner-$3j$ symbols,
\ba
\mathcal{G}^{\ell L L_1}_{mMM_1}
&=
\(\frac{(2\ell+1)(2L+1)(2L_1+1)}{4\pi}\)^\frac12
\begin{pmatrix}
  \ell & L & L_1 \\
  0 & 0 & 0
\end{pmatrix}
\begin{pmatrix}
  \ell & L & L_1 \\
  m & M & M_1
\end{pmatrix}\,.
\label{eq:gaunt_3j}
\ea
Hence, a product of two spherical harmonics can be reduced to a linear combination of spherical harmonics by
\ba
Y_{l_1m_1}(\rhat)Y_{l_2m_2}(\rhat)=\sum_{L}(-1)^M \mathcal{G}_{m_1m_2-M}^{l_1l_2L}Y_{LM}(\rhat)\,,
\ea
where $M=m_1 + m_2$.
Using this identity, one can derive by recursion the analogous integral to Eq.~\ref{eq:gaunt_factor} for any number of spherical harmonics. For four spherical harmonics we have (as in Appendix A of Ref.~\cite{slepian2018decoupling}):
\ba
&\int\dd^2\rhat \ Y_{l_1 m_1}(\rhat) Y_{l_2 m_2}(\rhat) Y_{l_3 m_3}(\rhat) Y_{l_4 m_4}(\rhat)= \sum_{L} (-1)^M\mathcal{G}^{l_1 l_2 L}_{m_1 m_2 -M} \mathcal{G}^{L l_3 l_4}_{M m_3 m_4}\,.
\label{eq:h_defn}
\ea
\subsection{Wigner symbols}
The Wigner $3j$ symbols obey an orthogonality relation
\ba
\sum_{mM}
\begin{pmatrix}
  \ell & L & L_1 \\
  m & M & M_1
\end{pmatrix}
\begin{pmatrix}
  \ell & L & L_2 \\
  m & M & M_2
\end{pmatrix}
&=
\frac{\delta^K_{L_1L_2}
\delta^K_{M_1M_2}
\delta^T(\ell,L,L_1)}
{2L_1+1}\,,
\label{eq:gaunt_orthogonality}
\ea
where $\delta^T(\ell,L,L_1)$ enforces the triangle relation that is obeyed
by the Wigner 3$j$-symbols, i.e they vanish unless $|\ell - L| \leq L_1 \leq \ell + L$ and $m + M + M_1 = 0$.

The Wigner $3j$'s acquire a phase for $m_i\rightarrow -m_i$:
\ba
\begin{pmatrix}
  j_1 & j_2 & j_3\\
  -m_1 & -m_2 & -m_3
\end{pmatrix}
=
(-1)^{j_1+j_2+j_3}
\begin{pmatrix}
  j_1 & j_2 & j_3\\
  m_1 & m_2 & m_3
\end{pmatrix}\,.
\label{eq:wigner3j_time_reversal}
\ea
We also have the identity
\begin{align}
    \begin{pmatrix}j_{1}&j_{2}&j_{3}\\
m_{1} & m_{2} & m_{3}\end{pmatrix}\begin{Bmatrix}j_{1} &j_{2} & j_{3}\\
l_{1} &l_{2} &l_{3}\end{Bmatrix}=\sum_{m^{\prime}_{1}m^{\prime}_{2}m^{\prime}_{3}}(-1)^{l_{1}+l_{2}+l_{3}+m^{\prime}_{1}+m^{\prime}_{2}+m^{\prime}_{3}}\begin{pmatrix}j_{1}&l_{2}&l_{3}\\
m_{1}&m^{\prime}_{2}&-m^{\prime}_{3}\end{pmatrix}\begin{pmatrix}l_{1}&j_{2}&l_{3}\\
-m^{\prime}_{1}&m_{2}&m^{\prime}_{3}\end{pmatrix}\begin{pmatrix}l_{1}&l_{2}&j_{3}\\
m^{\prime}_{1}&-m^{\prime}_{2}&m_{3}\end{pmatrix}
\label{eq:3jto6j} \,,
\end{align}
where a Wigner $6j$-symbol appears on the LHS. Lastly, we also have
\ba
\begin{pmatrix}j_{13}&j_{23}&j_{33}\\
m_{13}&m_{23}&m_{33}\end{pmatrix}\begin{Bmatrix}j_{11}&j_{12}&j_{13}\\
j_{21}&j_{22}&j_{23}\\
j_{31}&j_{32}&j_{33}\end{Bmatrix}
=\sum_{m_{r1},m_{r2},r=1,2,3}\begin{pmatrix}j%
_{11}&j_{12}&j_{13}\\
m_{11}&m_{12}&m_{13}\end{pmatrix}\begin{pmatrix}j_{21}&j_{22}&j_{23}\\
m_{21}&m_{22}&m_{23}\end{pmatrix} 
\begin{pmatrix}j_{31}&j_{32}&j_{33}\\
m_{31}&m_{32}&m_{33}\end{pmatrix}\nonumber \\
\times
\begin{pmatrix}j_{11}&j_{21}&j_{31}\\
m_{11}&m_{21}&m_{31}\end{pmatrix} 
\begin{pmatrix}j_{12}&j_{22}&j_{32}\\
m_{12}&m_{22}&m_{32}
\end{pmatrix}\,,
\label{eq:9j}
\ea
where a Wigner $9j$-symbol appears on the LHS.

\clearpage
\section{Encoding of observational isotropy by the Gaunt factor}
\label{app:isotropy}
Here we show that the $3$-point function of the SFB modes of an observationally-isotropic real-valued field $\delta(\mathbf{r})$ is real and proportional to the Gaunt factor. To see this, note that in real space, the 3-point function of $\delta$ can only depend on the distances to each point and the angles on the sky. Therefore, we may expand it in Legendre polynomials as
\ba
    \langle\delta(\bm{r}_1) \delta(\bm{r}_2) \delta(\bm{r}_3)\rangle
    &=
    \sum_{L_1 L_2 L_3} f_{L_1 L_2 L_3}(r_1, r_2, r_3)
    \,\mathcal{L}_{L_1}(\hat{\bm{r}}_1\cdot\hat{\bm{r}}_2)
    \,\mathcal{L}_{L_2}(\hat{\bm{r}}_2\cdot\hat{\bm{r}}_3)
    \,\mathcal{L}_{L_3}(\hat{\bm{r}}_3\cdot\hat{\bm{r}}_1)
    \label{eq:3point_rspace}
\ea
The Legendre polynomials may be further decomposed into sums over spherical harmonics via Eq.~\ref{eq:legendre_spherical_harmonics}. We may then transform Eq.~\ref{eq:3point_rspace} to spherical harmonic space to obtain
\begin{align}
    &\langle\delta_{\ell_1 m_1}(r_1) \delta_{\ell_2 m_2}(r_2) \delta_{\ell_3 m_3}(r_3)\rangle
    \nonumber\\
    &=
    \int d^2\hat{r}_1\,d^2\hat{r}_2\,d^2\hat{r}_3
    \,Y^*_{\ell_1 m_1}(\hat{\bm{r}}_1)
    \,Y^*_{\ell_2 m_2}(\hat{\bm{r}}_2)
    \,Y^*_{\ell_3 m_3}(\hat{\bm{r}}_3)
    \nonumber\\&\quad\times
    \sum_{L_1 L_2 L_3} A_{L_1L_2L_3} \, f_{L_1 L_2 L_3}(r_1, r_2, r_3)
    \sum_{M_1 M_2 M_3} (-1)^{M_1 + M_2 + M_3}
    \,Y^*_{L_1 -M_1}(\hat{\bm{r}}_1)Y^*_{L_1 M_1}(\hat{\bm{r}}_2)
    \nonumber\\&\quad\times
    \,Y^*_{L_2 -M_2}(\hat{\bm{r}}_2)Y^*_{L_2 M_2}(\hat{\bm{r}}_3)
    \,Y^*_{L_3 -M_3}(\hat{\bm{r}}_3)Y^*_{L_3 M_3}(\hat{\bm{r}}_1)   \nonumber \\
    &=
    \sum_{L_1 L_2 L_3} A_{L_1L_2L_3} \,f_{L_1 L_2 L_3}(r_1, r_2, r_3)
    \sum_{M_1 M_2 M_3} (-1)^{M_1 + M_2 + M_3}
    \,\mathcal{G}^{\ell_1 L_1 L_3}_{m_1 -M_1 M_3}
    \,\mathcal{G}^{\ell_2 L_2 L_1}_{m_2 -M_2 M_1}
    \,\mathcal{G}^{\ell_3 L_3 L_2}_{m_3 -M_3 M_2}\,.
\end{align}
where we define $A_{L_1L_2L_3} \equiv \prod_i 4\pi/(2L_i+1)$. Using the identity Eq.~\ref{eq:3jto6j} to evaluate the sum of Gaunt factors, we may write
\ba
\langle\delta_{\ell_1 m_1}(r_1) \delta_{\ell_2 m_2}(r_2) \delta_{\ell_3 m_3}(r_3)\rangle&=\sqrt{(4\pi)^3(2l_1+1)(2l_2+1)(2l_3+1)}\begin{pmatrix}l_{1}&l_{2}&l_{3} \nonumber\\
m_{1} & m_{2} & m_{3}\end{pmatrix} \nonumber\\ 
\times \sum_{L_1 L_2 L_3} f_{L_1 L_2 L_3}(r_1, r_2, r_3)&(-1)^{L_1+L_2+L_3}\begin{Bmatrix}l_{1} &l_{2} & l_{3}\\
L_{1} &L_{2} &L_{3}\end{Bmatrix}\begin{pmatrix}l_{1}&L_{1}&L_{3}\\
0 & 0 & 0\end{pmatrix}\begin{pmatrix}l_{2}&L_{2}&L_{1} \\
0 & 0 & 0\end{pmatrix}\begin{pmatrix}l_{3}&L_{3}&L_{2}\\
0 & 0 & 0\end{pmatrix}.
\ea
The Wigner $3j$ symbols inside the sum over the $L_i$ impose that $l_1+l_2+l_3$ be even, hence the sum is proportional to the symbol $\begin{pmatrix}l_{1}&l_{2}&l_{3}\\
0 & 0 & 0\end{pmatrix}$
, and $\langle\delta_{\ell_1 m_1}(r_1) \delta_{\ell_2 m_2}(r_2) \delta_{\ell_3 m_3}(r_3)\rangle$, to the Gaunt factor $\mathcal{G}^{\l_1 l_2 l_3}_{m_1 m_2 m_3}$, which encodes the isotropy. The $3$-point function of the SFB modes is then obtained by applying the basis transformation Eq.~\ref{eq:sfb_fourier_pair_b},
hence it is real.
\clearpage
\section{An identity for integrating a product of three spherical harmonics }
\label{app:derivation_identity}

Here we derive the identity 
\ba
I_{m_1m_2m_3}^{l_1l_2l_3}(k_1,k_2,k_3)&\equiv\int d^2\khat_1d^2 \khat_2d^2\khat_3Y^*_{l_1,m_1}(\khat_1)Y^*_{l_2,m_2}(\khat_2)Y^*_{l_3,m_3}(\khat_3)\delta_D(\vk_1+\vk_2+\vk_3) \nonumber \\
&=  
 \frac{4\pi^{\frac{3}{2}}}{k_1k_2k_3}\sqrt{(2l_1+1)}
\begin{pmatrix}
  l_1 & l_2 &  l_3 \\
  m_1 & m_2 & m_3
\end{pmatrix}  \sum_{|m| \le \min(l_2,l_3)} \!\!\!\!\!{Y_{l_2,m}}(\theta_{12},0)Y_{l_3,-m}(\theta_{13},0)
(-1)^m
\begin{pmatrix}
  l_1 & l_2 & l_3 \\
  0 & m & -m
\end{pmatrix}
\,,
\label{eq:gauntlike_identity_app}
\ea
which we use to calculate the SFB bispectrum in the homogeneous and isotropic Universe, as well as to accelerate part of the calculation of the observed bispectrum.

We need only to show this for $k_1,k_2,k_3$ which satisfy the triangle inequality, as otherwise the integral clearly vanishes. Above, we define $\theta_{ij}$ as the angle between $\vk_i$ and $\vk_j$, such that $\cos(\theta_{12}) = \vartheta(k_1,k_2,k_3) \equiv\frac{k_3^2-k_1^2-k_2^2}{2k_1k_2}$ and $\cos({\theta_{13}}) = \vartheta(k_1,k_3,k_2)$, and we denote for the unit vector $\rhat$ of spherical angles $(\theta,\phi)$, $Y_{\ell,m}(\theta,\phi)\equiv Y_{\ell,m}(\rhat)$. The angle-averaged form of Eq.~\ref{eq:gauntlike_identity_app} is denoted $I_{l_1l_2l_3}(k_1,k_2,k_3)$ such that 
\ba
I_{m_1m_2m_3}^{l_1l_2l_3}(k_1,k_2,k_3)=\begin{pmatrix}
  l_1 & l_2 &  l_3 \\
  m_1 & m_2 & m_3
\end{pmatrix} I_{l_1l_2l_3}(k_1,k_2,k_3) \,.
\label{eq:Il1l2l3} 
\ea 

We begin by using Eq.~\ref{eq:dirac_composition} and Eq.~\ref{eq:dirac3D}  to write
\ba
\delta_D(\vk_1+\vk_2+\vk_3)&=k_3^{-2}\delta_D(|\vk_1+\vk_2|-k_3)\delta_D(\widehat{\vk_1+\vk_2}+\khat_3) \nonumber \\
&=(k_1k_2k_3)^{-1}\delta_D(\khat_1\cdot\khat_2-\vartheta(k_1,k_2,k_3))\delta_D(\widehat{\vk_1+\vk_2}+\khat_3) \,,
\label{eq:dirac_sum3}
\ea
such that, after integration over $d^2\khat_3$, Eq.~\ref{eq:gauntlike_identity_app} becomes
\ba
I_{m_1m_2m_3}^{l_1l_2l_3}(k_1,k_2,k_3)=(k_1k_2k_3)^{-1}(-1)^{l_3}\int d^2\khat_1d^2\khat_2
Y^*_{l_1,m_1}(\khat_1)Y^*_{l_2,m_2}(\khat_2)Y^*_{l_3,m_3}((\widehat{\vk_1+\vk_2}))\delta_D(\khat_1\cdot\khat_2-\vartheta(k_1,k_2,k_3))
\label{eq:3point_intk1k2} \,,
\ea
where we used the parity property Eq.~\ref{eq:Ylm_parity}. We then integrate over $\khat_2$ by rotating it through an angle $\varphi_{2}$ around $\khat_1$, as the Dirac delta fixes $\cos(\theta_{12}) =\vartheta(k_1,k_2,k_3)$. Then $\cos(\pi - \theta_{13})$ also remains fixed and $\widehat{\vk_1+\vk_2}$ rotates about $\khat_1$ by the same angle $\varphi_2$. 

We denote by $\mathcal{R}(\khat_1)$ the rotation sending the axis $\mathbf{\hat{z}}$ to $\khat_1$. Using the rotation formula for spherical harmonics Eq.~\ref{eq:rotation_ylm} and integrating over $\khat_1 \cdot \khat_2$, Eq.~\ref{eq:3point_intk1k2} becomes 
\ba
I_{m_1m_2m_3}^{l_1l_2l_3}(k_1,k_2,k_3)=(k_1k_2k_3)^{-1}(-1)^{l_3}\int d^2\khat_1d\varphi_2 Y^*_{l_1,m_1}&(\khat_1)\bigg[\sum_{m_2' = -l_2}^{l_2} D^{(l_2)}_{m_2,m_2'}({\mathcal{R}(\khat_1)})Y^*_{l_2,m_2'}(\theta_{12},\varphi_2)\bigg] \nonumber \\
&\times \bigg[\sum_{m_3' = -l_3}^{l_3} D^{(l_3)}_{m_3,m_3'}({\mathcal{R}(\khat_1)}) Y^*_{l_3,m_3'}(\pi-\theta_{13},\varphi_2)\bigg] \,,
\ea
where $D^{(\ell)}_{m,m'}(\mathcal{R}(\khat_1))$ are Wigner $D$-matrix elements. As $Y^*_{\ell,m}(\theta,\phi)$ is proportional to $e^{-im\phi}$, integrating the pairwise products of spherical harmonics over $d \varphi_2$ gives factors $2\pi \delta^K_{m_2',-m_3'}$. Hence
\ba
I_{m_1m_2m_3}^{l_1l_2l_3}(k_1,k_2,k_3)=2\pi(k_1k_2k_3)^{-1}(-1)^{l_3}\sum_{m_2'=-\min(l_2,l_3)}^{\min(l_2,l_3)}&Y^*_{l_2,m_2'}(\theta_{12},0)Y^*_{l_3,-m_2'}(\pi-\theta_{13},0) \nonumber \\
&\times \int d^2\khat_1 Y^*_{l_1,m_1}(\khat_1) D^{(l_2)}_{m_2,m_2'}({\mathcal{R}(\khat_1)})D^{(l_3)}_{m_3,-m_2'}({\mathcal{R}(\khat_1)})
\label{eq:Ik1k2k3} \,,
\ea
where $D^{(\ell)}_{mm'}({\mathcal{R}})$ is the Wigner $D$-matrix.
Using Eq.~\ref{eq:WignerD_pairwise_prod}, the integral in Eq.~\ref{eq:Ik1k2k3} becomes 
\ba
\int d^2\khat_1 Y^*_{l_1,m_1}(\khat_1) D^{(l_2)}_{m_2,m_2'}({\mathcal{R}(\khat_1)})D^{(l_3)}_{m_3,-m_2'}({\mathcal{R}(\khat_1)})&=
 \sum_{J=|l_2-l_3|}^{l_2+l_3}  \langle l_2 m_2 l_3 m_3 | J \left(m_2 + m_3\right) \rangle
               \langle l_2 m_2' l_3 (-m_2') | J 0 \rangle \nonumber          \\
               &\times \int d^2\khat_1 Y^*_{l_1,m_1}(\khat_1)
  D^J_{m_2 + m_3,0}({\mathcal{R}(\khat_1)}) \nonumber\\
  &=
 (-1)^{m_1}\sqrt{\frac{4\pi}{2l_1+1}}\langle l_2 m_2 l_3 m_3 | l_1 \left(-m_1\right) \rangle
               \langle l_2 m_2' l_3 (-m_2') | l_1 0 \rangle \,,
               \label{eq:intermediate3harmonic_integral}
\ea
where the second equation above follows from the first by applying the identity Eq.~\ref{eq:WignerD_lm0} to write
\ba
D^J_{m_2 + m_3,0}({\mathcal{R}(\khat_1)})=\sqrt{\frac{4\pi}{2J+1}}Y^*_{J,m_2+m_3}(\khat_1) \,,
\ea
and then using the orthogonality of spherical harmonics Eq.~\ref{eq:YlmYlmDelta}. We obtain the final expression Eq.~\ref{eq:gauntlike_identity_app} by expressing the Clebsch-Gordan coefficients in terms of Wigner $3j$'s via Eq.~\ref{eq:WignerD_toWigner3j} and inserting Eq.~\ref{eq:intermediate3harmonic_integral} into Eq.~\ref{eq:Ik1k2k3}. In writing Eq.~\ref{eq:gauntlike_identity_app} we have also removed the complex conjugations from the spherical harmonics as they are real, and used the parity of associated Legendre polynomials Eq.~\ref{eq:plm_parity}.  

It is also instructive to rewrite the integral Eq.~\ref{eq:gauntlike_identity_app} to make explicit the consequence of isotropy. Using that
\ba
\delta_D(\vk_1+\vk_2+\vk_3) = \frac{1}{(2\pi)^3}\int \dd^3 \vr e^{i(\vk_1+\vk_2+\vk_3)\cdot \vr}
\label{eq:diracsum} \,,
\ea
and Eq.~\ref{eq:spherical_transform_g1} (for $\alpha=0$),
Eq.~\ref{eq:gauntlike_identity_app} becomes
\ba
I_{m_1m_2m_3}^{l_1l_2l_3}(k_1,k_2,k_3)=8
i^{l_1+l_2+l_3}\int r^2\dd r j_{l_1}(k_1r)j_{l_2}(k_2r)j_{l_3}(k_3r)\int d^2\rhat Y^*_{l_1,m_1}(\rhat)Y^*_{l_2,m_2}(\rhat)Y^*_{l_3,m_3}(\rhat) \,,
\label{eq:gauntlike_propto_gaunt}
\ea
where the angular integral is the Gaunt factor $\mathcal{G}^{l_1l_2l_3}_{m_1m_2m_3}$ encoding the isotropy. The above radial integral of the product of three spherical Bessel functions has been evaluated analytically and by recursion in \cite{Mehrem_1991, fabrikant, dong_kim_chirikjian_2015, Chellino_2023}, though typically with methods requiring more computation than Eq.~\ref{eq:gauntlike_identity_app}. During the writing of this manuscript the authors discovered Ref.~\cite{maximonjackson}, which also evaluated the radial integral, leading to a result equivalent to Eq.~\ref{eq:gauntlike_identity_app} via an alternate derivation.
\clearpage
\section{Details of the bispectrum computation}
\label{app:bispectrum_comutation_details}

\subsection{Legendre expansion coefficients $Z_{l_1l_2l_3}(r,k_1,k_2)$}
\label{app:Zlll_terms}

In this subsection we derive the coefficients $Z_{l_1l_2l_3}(r,k_1,k_2)$ introduced in Eq.~\ref{eq:z2_decomposed}, which we reproduce here:
\ba
\tilde{Z}_2(\vk_1,\vk_2,\vr)
&=
\sum_{l_1l_2l_3} Z_{l_1 l_2 l_3}(k_1,k_2,r)
\,\mathcal{L}_{l_1}(\khat_1\cdot\khat_2)
\,\mathcal{L}_{l_2}(\khat_1\cdot\rhat)
\,\mathcal{L}_{l_3}(\khat_2\cdot\rhat)
\vs
&=
\sum_{\{l_i m_i\}}
\frac{(4\pi)^3\,Z_{l_1 l_2 l_3}(k_1,k_2,r)}{\(2l_1 + 1\)\(2l_2 + 1\)\(2l_3 + 1\)}
\,Y_{l_1 m_1}(\khat_1)\,Y^*_{l_1 m_1}(\khat_2)
\,Y_{l_2 m_2}(\khat_1)\,Y^*_{l_2 m_2}(\rhat)
\,Y_{l_3 m_3}(\khat_2)\,Y^*_{l_3 m_3}(\rhat)
\,.
\label{eq:z2_decomposed_reproduced}
\ea
We first note that the form of $Z_2(\vk_1,\vk_2,\vr)$ (see Eq.~\ref{eq:z2_with_fnl}) implies that it 
can be written as a polynomial in $\mu_1$, $\mu_2$, $\khat_1\cdot\khat_2$, and $\mu=(k_1\mu_1+k_2\mu_2)/k$, 
except for the terms proportional to $f_\mathrm{NL}$:
\ba
f_{\mathrm{NL}} \frac{\alpha(k)}{\alpha\left(k_{1}\right) \alpha\left(k_{2}\right)}\left(b^\mathrm{E}_{10} + f(r)\mu^2\right)
\label{eq:fnl_term} \,.
\ea
Furthermore, except for the term
\ba
f(r) \mu^{2}G_{2}\left(\vk_{1}, \vk_{2}\right)
\label{eq:G2_term} \,,
\ea
$Z_2(\vk_1,\vk_2,\vr)$ is a polynomial of in $\mu_1$, $\mu_2$ and $\khat_1\cdot\khat_2$ alone, which allows us to write the decomposition Eq.~\ref{eq:z2_decomposed}. In fact, Eq.~\ref{eq:G2_term} could also be included in this decomposition, by writing
\ba
\mu^2=\frac{(\mu k)^2}{k_1^2+k_2^2+2k_1k_2\khat_1\cdot\khat_2} = \frac{(k_1\mu_1+k_2\mu_2)^2}{k_1^2+k_2^2}\sum_{n \ge 0}\bigg(-\frac{2k_1k_2}{k_1^2+k_2^2}\khat_1\cdot\khat_2\bigg)^n \,.
\label{eq:mu2_decomposed}
\ea
However, then the sum over $l_1,l_2,l_3$ in Eq.~\ref{eq:z2_decomposed} becomes infinite, and truncating this slowly-converging series after even a small number of terms greatly increases the number of non-vanishing coefficients $Z_{l_1l_2l_3}$. Hence, we opt to treat Eq.~\ref{eq:G2_term} along with  Eq.~\ref{eq:fnl_term} separately (and exactly) as described in Appendix \ref{app:velocity_div_kernel}. 

Leaving out these terms, we have only $9$ coefficients $Z_{l_1l_2l_3}(r,k_1,k_2)$, which are listed below by triplet $(l_1,l_2,l_3)$. We have dropped the implicit $r$-dependence for brevity. 
\ba
 (0, 0, 0)  &: \quad  \frac{1}{3}b^\mathrm{E}_{10}f + \frac{17}{21}b^\mathrm{E}_{10} + \frac{1}{2}b^\mathrm{E}_{20} + \frac{1}{9}f^2 +\frac{b^\mathrm{E}_{01}f}{6\alpha(k_1)}+\frac{b^\mathrm{E}_{01}f}{6\alpha(k_2)}+\frac{b^\mathrm{E}_{11}}{2\alpha(k_1)}+\frac{b^\mathrm{E}_{11}}{2\alpha(k_2)}+\frac{b^\mathrm{E}_{02}}{\alpha(k_1)\alpha(k_2)} \\
 (0, 0, 2) &: \quad  \frac{1}{9}f(3b^\mathrm{E}_{10} + 2f) +f\frac{b^\mathrm{E}_{01}}{3\alpha(k_1)}   \\
 (0, 1, 1) &: \quad  \bigg(\frac{k_1}{2k_2}+\frac{k_2}{2k_1}\bigg)f(b^\mathrm{E}_{10}+\frac{3}{5}f)    +b^\mathrm{E}_{01}f \bigg(\frac{k_2}{2k_1\alpha(k_2)}+\frac{k_1}{2k_2\alpha(k_1)}\bigg) \\
  (0, 1, 3) &: \quad  f^2 \frac{k_2}{5k_1}     \\
 (0, 2, 0)&: \quad  \frac{1}{9}f(3b^\mathrm{E}_{10} + 2f) +f\frac{b^\mathrm{E}_{01}}{3\alpha(k_2)}   \\
 (0, 2, 2)&: \quad  \frac{4}{9}f^2 \\
 (0, 3, 1)&: \quad  f^2 \frac{k_1}{5k_2}     \\
 (1, 0, 0)&: \quad  b^\mathrm{E}_{10}\bigg(\frac{k_1}{2k_2}+\frac{k_2}{2k_1}\bigg) +b^\mathrm{E}_{01}\bigg(\frac{k_2}{2k_1\alpha(k_2)}+\frac{k_1}{2k_2\alpha(k_1)}\bigg) \\
 (2, 0, 0)&: \quad \frac{4}{21}
 \ea

\clearpage
\subsection{Derivation of the observed SFB bispectrum Eq.~\ref{eq:sfb_tot}} 
Here we detail the remainder of the derivation of the observed bispectrum Eq.~\ref{eq:sfb_tot} after the Legendre expansion of the second-order redshift-space kernel has been performed as in Section~\ref{app:Zlll_terms}.

We first insert the decomposition of $\tilde{Z}_2(\vk_1,\vk_2,\vr)$ (Eq.~\ref{eq:z2_decomposed_reproduced}) into $\delta_{lm}^{g,(2)}$ (Eq.~\ref{eq:delta_2_expanded}), and use the identity in Eq.~\ref{eq:rayleigh} to simplify the complex exponentials. We rid of the angular integrals over $\qhat$ with the orthogonality relation for spherical harmonics Eq.~\ref{eq:YlmYlmDelta}. After
rearranging for the angular and radial integrals and assuming a separable window $W(\vr)=W(r)W(\rhat)$, we obtain
\ba
\delta_{\ell m}^{g,(2)}(k)
&=
\sqrt{\frac{2}{\pi}}\,k
\int\dd r\,r^2
\,j_\ell(kr)
\,W(r) \,D^2(r)
\,4\pi
\int\frac{\dd k_1\,k_1}{(2\pi)^3}
\int\frac{\dd k_2\,k_2}{(2\pi)^3}
\sum_{\{l_im_i\}}
\frac{(4\pi)^3\,Z_{l_1 l_2 l_3}(k_1,k_2,r)}{\(2l_1 + 1\)\(2l_2 + 1\)\(2l_3 + 1\)}
\vs&\quad\times
r^2
\,(4\pi)^3
\sum_{\{L_iM_i\}}
\sum_{LM}
i^{-L_1 -L_2 +L_3+L_4}
\,j_{L_3}(k_1r)
\,j_{L_4}(k_2r)
\,\delta^{(1)}_{L_1 M_1}(k_1)
\,\delta^{(1)}_{L_2 M_2}(k_2)
\vs&\quad\times
\frac{\pi}{2r^2}
\,\Gaunt{L_3}{L_4}{L}{M_3}{M_4}{M}
\int\dd^2\rhat\,
W(\rhat)
\,Y_{LM}(\rhat)
\,Y^*_{\ell m}(\rhat)
\,Y^*_{l_2 m_2}(\rhat)
\,Y^*_{l_3 m_3}(\rhat)
\vs&\quad\times
\int\dd^2\khat_1
\,Y_{l_1 m_1}(\khat_1)
\,Y_{l_2 m_2}(\khat_1)
\,Y_{L_3M_3}(\khat_1)
\,Y_{L_1 M_1}(\khat_1)
\int\dd^2\khat_2
\,Y^*_{l_1 m_1}(\khat_2)
\,Y_{l_3 m_3}(\khat_2)
\,Y_{L_4M_4}(\khat_2)
\,Y_{L_2 M_2}(\khat_2)
\,.
\ea
We may rewrite this more compactly as
\ba
\delta_{\ell m}^{g,(2)}(k)
&=
\int\dd k_1
\int\dd k_2
\sum_{L_1M_1}
\sum_{L_2M_2}
\mathcal{V}^{\ell L_1 L_2 }_{m M_1 M_2}(k,k_1,k_2)
\,\delta^{(1)}_{L_1 M_1}(k_1)
\,\delta^{(1)}_{L_2 M_2}(k_2)
\,,
\label{eq:delta2_compactly}
\ea
where we use $\mathcal{V}$ to denote the second-order coupling kernel in the SFB bispectrum 
\ba
\mathcal{V}^{\ell L_1 L_2}_{m M_1 M_2}(k,k_1,k_2)
&\equiv
\sqrt{\frac{\pi}{2}}
\,2^8\pi
\,kk_1k_2
\int\dd r\,r^2
\,j_\ell(kr)
\,W(r) \,D^2(r)
\sum_{l_1 l_2 l_3}
\frac{Z_{l_1 l_2 l_3}(k_1,k_2,r)}{\(2l_1 + 1\)\(2l_2 + 1\)\(2l_3 + 1\)}
\vs&\quad\times
\sum_{L_3 L_4}
\,j_{L_3}(k_1r)
\,j_{L_4}(k_2r)
\,\mathcal{C}^{L_1}_{M_1}{}^{L_2}_{M_2}{}^{\ell}_{m}{}^{,L_3 L_4}_{,l_1 l_2 l_3}\,.
\label{eq:wlll}
\ea
Here ${C}^{L_1}_{M_1}{}^{L_2}_{M_2}{}^{\ell}_{m}{}^{,L_3 L_4}_{,l_1 l_2 l_3}$ is a mode coupling coefficient
\ba
\mathcal{C}^{L_1}_{M_1}{}^{L_2}_{M_2}{}^{\ell}_{m}{}^{,L_3 L_4}_{,l_1 l_2 l_3}
&=
i^{-L_1 -L_2 +L_3+L_4}\sum_{m_1 m_2 m_3}
\sum_{M_3 M_4}\sum_{LM} (-1)^{M+m_1}
\mathcal{G}^{L_3L_4L}_{M_3M_4M}
\mathcal{H}^{L\ell l_2 l_3}_{-M m m_2 m_3}
\mathcal{H}^{l_1l_2L_3L_1}_{m_1m_2M_3M_1}
\mathcal{H}^{l_1l_3L_4L_2}_{-m_1m_3M_4M_2}\,.
\label{eq:coeff_C}
\ea
and  
\ba
&\mathcal{H}^{l_1 l_2 l_3 l_4}_{m_1 m_2 m_3 m_4} \equiv \int\dd^2\rhat \ W(\rhat) Y_{l_1 m_1}(\rhat) Y_{l_2 m_2}(\rhat) Y_{l_3 m_3}(\rhat) Y_{l_4 m_4}(\rhat)\,.
\label{eq:4ylmwindow}
\ea
is the integral over four spherical harmonics and the angular part of the window function.

Using the form of $\delta^{g,(1)}_{\ell m}(k)$ in Eq.~\ref{eq:sfb_coeff_obs}, the terms contributing to the tree-level $3$-point correlation function of the SFB modes are
\ba
&\<
\delta_{\ell m}^{g,(2)}(k)
\,\delta^{g,(1)}_{\ell' m'}(k')
\,\delta^{g,(1)}_{\ell'' m''}(k'')
\>
\vs
&=
\int\dd k_1
\int\dd k_2
\sum_{L_1M_1}
\sum_{L_2M_2}
\mathcal{V}^{\ell L_1 L_2 }_{m M_1 M_2}(k,k_1,k_2)
\int\dd q'
\sum_{L'M'}
\mathcal{W}_{\ell' m'}^{L'M'}(k',q')
\int\dd q''
\sum_{L''M''}
\mathcal{W}_{\ell'' m''}^{L''M''}(k'',q'')
\vs
&\quad\times
\<\delta^{(1)}_{L_1 M_1}(k_1)
\,\delta^{(1)}_{L_2 M_2}(k_2)
\,\delta^{(1)}_{L'M'}(q')
\,\delta^{(1)}_{L''M''}(q'')\>\,,
\ea
along with the two other terms with cyclically permuted superscript indices.
Noting that the SFB power spectrum for the
constant-time slice matter density contrast is homogeneous and isotropic, i.e.
\ba
\<\delta^{(1)}_{L_1 M_1}(k_1)
\,\delta^{(1)}_{L_2 M_2}(k_2)\>
&= \delta^K_{L_1 L_2} \delta^K_{M_1 -M_2}(-1)^{M_2} \delta^D(k_1-k_2)\,P(k_1)\,,
\ea
we may apply Wick's theorem,
\ba
&\<
\delta_{\ell m}^{g,(2)}(k)
\,\delta^{g,(1)}_{\ell' m'}(k')
\,\delta^{g,(1)}_{\ell'' m''}(k'')
\>
\vs
&=
\int\dd k_1
\sum_{L_1M_1}
(-1)^{M_1}\mathcal{V}^{\ell L_1 L_1}_{m M_1 -M_1}(k,k_1,k_1)
\,P(k_1)
\int\dd q'
\sum_{L'M'}(-1)^{M'}
\mathcal{W}_{\ell' m'}^{L'M'}(k',q')
\,\mathcal{W}_{\ell'' m''}^{L'-M'}(k'',q')
\,P(q')
\vs&\quad
+
\int\dd q'
\int\dd q''
\sum_{L'M'}
\sum_{L''M''}
(-1)^{M'+M''}\mathcal{V}^{\ell L' L'' }_{m -M' -M''}(k,q',q'')
\,\mathcal{W}_{\ell' m'}^{L' M'}(k',q')
\,\mathcal{W}_{\ell'' m''}^{L''M''}(k'',q'')
\,P(q')\,P(q'')
\vs&\quad
+
\int\dd q'
\int\dd q''
\sum_{L''M''}
\sum_{L'M'}
(-1)^{M'+M''}\mathcal{V}^{\ell L'' L' }_{m -M'' -M'}(k,q'',q')
\,\mathcal{W}_{\ell' m'}^{L'M'}(k',q')
\,\mathcal{W}_{\ell'' m''}^{L''M''}(k'',q'')
\,P(q')\,P(q'')\,.
\ea

From here on we assume a spherically symmetric window $W(\vr)=W(r)$. Then we have $\mathcal{W}_{\ell m}^{LM}(k,q) = \delta^K_{\ell L} \delta^K_{mM}\,\mathcal{W}_\ell(k,q)$. Hence,
\ba
&\<
\delta_{\ell m}^{g,(2)}(k)
\,\delta^{g,(1)}_{\ell' m'}(k')
\,\delta^{g,(1)}_{\ell'' m''}(k'')
\>
\vs
&=
\delta^K_{\ell'\ell''}
\delta^K_{m'-m''}(-1)^{m'}
\,C_{\ell'}(k',k'')
\int\dd k_1
\sum_{L_1M_1}
(-1)^{M_1}\mathcal{V}^{\ell L_1 L_1 }_{m M_1 -M_1}(k,k_1,k_1)
\,P(k_1)
\vs&\quad
+
\int\dd q'
\,\mathcal{W}_{\ell'}(k',q')
\,P(q')
\int\dd q''
\,\mathcal{W}_{\ell''}(k'',q'')
\,P(q'')(-1)^{m'+m''}
\left[\mathcal{V}^{\ell \ell' \ell''}_{m -m' -m''}(k,q',q'')
+\mathcal{V}^{\ell \ell'' \ell'}_{m -m'' -m'}(k,q'',q')\right].
\label{eq:3point_211}
\ea

In Eq.~\ref{eq:delta2_compactly} swapping $L_1 \leftrightarrow L_2, k_1 \leftrightarrow k_2,M_1 \leftrightarrow M_2$ gives $\mathcal{V}^{\ell L_1 L_2 }_{m M_1 M_2}(k,k_1,k_2)=\mathcal{V}^{\ell L_2 L_1 }_{m M_2 M_1}(k,k_2,k_1)$, thus the two terms in brackets in Eq.~\ref{eq:3point_211} are equal. In Appendix \ref{app:wll_simplified} we simplify $\mathcal{V}^{\ell \ell' \ell''  }_{m -m'-m''}$ and show that it is proportional to the Gaunt factor $\mathcal{G}^{\ell \ell' \ell''}_{m m'm''}$. Hence we define the angle-averaged quantity
\ba
\mathcal{V}^{\ell \ell' \ell''}(k,k',k'')\equiv
\sum_{mm'm''}
\begin{pmatrix}
 \ell && \ell' && \ell'' \\
 m && m' && m''
\end{pmatrix}
(-1)^{m'+m''}\mathcal{V}^{\ell \ell' \ell'' }_{m -m'-m''}(k,k',k'') \,,
\label{eq:wll_angle_averaged}
\ea
whose simplified expression is given by Eq.~\ref{eq:wlll_averaged}. 

In fact, the first line of Eq.~\ref{eq:3point_211} always vanishes when $\ell>0$, which we may see by the following quick argument. By the remarks at the beginning of Section~\ref{sec:SFBbispectrum_def} on observational isotropy,  Eq.~\ref{eq:3point_211} must be proportional to the Gaunt factor $\mathcal{G}^{\ell \ell' \ell''}_{m m'm''}$, such that Eq.~\ref{eq:3point_211} is nonzero only if $\ell,\ell',\ell''$ form a triangle. This condition is already imposed by the second line of Eq.~\ref{eq:3point_211}.  Let $\ell>0$ and set $\ell'=\ell''=0$ such that the triangle condition is violated. As $C_0(k',k'') \neq 0$, the  integral in the first line of Eq.~\ref{eq:3point_211} must vanish, and consequently the first line must vanish for any $(\ell,\ell',\ell'')$ with  $\ell>0$. The authors of Ref.~\cite{bertacca} demonstrate this directly with a lengthy derivation. 

Finally, after angle-averaging with Eq.~\ref{eq:sfb_3point_reduced} and re-indexing for clarity, the SFB bispectrum, ignoring contributions from the $f_\mathrm{NL}$ and $G_2$ terms, is given by 
\ba
B^{\text{SFB,}\,\tilde{Z}_2}_{l_1l_2l_3}(k_1,k_2,k_3)&=
2\int\dd q_2
\mathcal{W}_{l_2}(k_2,q_2)
\,P(q_2)
\int\dd q_3
\,\mathcal{W}_{l_3}(k_3,q_3)
\,P(q_3) \ \mathcal{V}^{l_1l_2l_3}(k_1,q_2,q_3)+ \mathrm{2\ cyc.\  perm. \,.}
\label{eq:sfb_bispec_final}
\ea
The contribution from the $f_{\mathrm{NL}}$ and $G_2$ terms is given in Appendix \ref{app:velocity_div_kernel} (Eq.~\ref{eq:sfb_G2_appendix}). It is of a similar form to the contribution from the terms in $\tilde{Z}_2$.  

In principle, it is possible to repeat the above derivation while relaxing the assumption that $W(\vr)$ is spherically symmetric (but keeping the assumption that the radial and angular dependencies are separable) by decomposing $W(\rhat)$ into spherical harmonics. However, as in this case we can no longer leverage observational isotropy of the signal, the calculation is significantly more complicated (and expensive), so we leave the details to a future work.

Lastly, it is useful to verify that we retrieve Eq.~\ref{eq:sfb_3point_expanded2} in the limit of an isotropic and homogeneous Universe, i.e by setting $D=b_1=W=1$ and $f=0$. In this case, the second term in Eq.~\ref{eq:sfb_tot} vanishes, and we may take $W_\ell(k,q)=\delta_D(k-q)$, such that
\ba
B^{\text{SFB}}_{l_1l_2l_3}(k_1,k_2,k_3)=2P(k_2)P(k_3) \ \mathcal{V}^{l_1l_2l_3}(k_1,k_2,k_3)+ \mathrm{2\ cyc.\  perm.} \,.
\label{eq:full_sfb_iso_homo_limit}
\ea
As it is somewhat tedious to demonstrate equivalence with Eq.~\ref{eq:sfb_3point_expanded2} analytically, we omit the details here\footnote{For a lengthy proof along these lines see Ref.~\cite{bertacca}.}; as a test of our code, we verify numerically that Eq.~\ref{eq:full_sfb_iso_homo_limit} and Eq.~\ref{eq:sfb_3point_expanded2} are identical in the considered limit. 
\subsection{Simplification of the kernel $\mathcal{V}^{L_1 L_2 \ell}_{M_1 M_2 m}(k,k_1,k_2)$}
\label{app:wll_simplified}
We now show that, under the assumption of a spherically symmetric window $W(\vr)=W(r)$, the kernel $\mathcal{V}^{\ell L_1 L_2 }_{m M_1 M_2}$ is proportional to the Gaunt factor $\mathcal{G}^{L_1 L_2 \ell}_{M_1 M_2 -m}$, and compute its angle-averaged expression, defined by Eq.~\ref{eq:wll_angle_averaged}.  For compactness, we will denote the Wigner coefficient $\begin{pmatrix} \ell_1 & \ell_2 & \ell_3 \\ m_1 & m_2 & m_3\end{pmatrix}$ by $K^{\ell_1\ell_2\ell_3}_{m_1m_2m_3}$
and the coefficient of proportionality between Gaunt factors and Wigner coefficients by
\ba
f_{\ell_1\ell_2\ell_3}
&\equiv
\(\frac{(2\ell_1+1)(2\ell_2+1)(2\ell_3+1)}{4\pi}\)^\frac12
\begin{pmatrix}
  \ell_1 & \ell_2 & \ell_3 \\
  0 & 0 & 0
\end{pmatrix}
\,,
\ea
such that $\mathcal{G}^{\ell_1\ell_2\ell_3}_{m_1m_2m_3}=f_{\ell_1\ell_2\ell_3}K^{\ell_1\ell_2\ell_3}_{m_1m_2m_3}$.

First, we use the identity Eq.~\ref{eq:h_defn} to evaluate the integral over four spherical harmonics Eq.~\ref{eq:4ylmwindow}, such that we may express Eq.~\ref{eq:coeff_C} in terms of Gaunt factors\footnote{Note that if we allow for a generic angular dependence of the window $W(\rhat)$, it is still possible to evaluate the integral Eq.~\ref{eq:h_defn} by decomposing $W(\rhat)$ into spherical harmonics and using the generalization of the integral Eq.~\ref{eq:coeff_C} to a product of five spherical harmonics. However this leads to an explosion in the number of terms needed to compute the kernel $\mathcal{V}^{\ell L_1 L_2 }_{m M_1 M_2}$, so in practice only spherically symmetric windows are currently computationally feasible.}. 
We also use that the sum of the lower indices of the Gaunt factor must vanish, to write
\begin{align}
\mathcal{C}^{L_1}_{M_1}{}^{L_2}_{M_2}{}^{\ell}_{m}{}^{,L_3 L_4}_{,l_1 l_2 l_3}
&=i^{-L_1 -L_2 +L_3+L_4}\sum_{L,H_1,H_2,H_3}\sum_{m_1,m_2,m_3,M_3,M_4,M}(-1)^{m+m_1+m_2+m_3} \nonumber \\
&\quad\times \mathcal{G}^{L_3L_4L}_{M_3M_4M}\mathcal{G}^{L\ell H_1}_{-M m -N_1} \mathcal{G}^{H_1 l_2 l_3}_{N_1 m_2 m_3}\mathcal{G}^{l_1l_2 H_2}_{m_1 m_2 -N_2} \mathcal{G}^{H_2 L_3 L_1}_{N_2 M_3 M_1}\mathcal{G}^{l_1 l_3 H_3}_{-m_1 m_3 -N_3} \mathcal{G}^{H_3 L_4 L_2}_{N_3 M_4 M_2}\,, \label{eq:B3}
\end{align}
where the sums over $N_1$, $N_2$, and $N_3$ have only one term each, and are thus not explicitly written.
With Eq.~\ref{eq:wigner3j_time_reversal}, and changing signs on the summation variables,
we can use Eq.~\ref{eq:3jto6j} to simplify the inner sum
\begin{align}
&\sum_{m_1m_2m_3}(-1)^{m_1+m_2+m_3} \mathcal{G}^{H_1 l_2 l_3}_{N_1 m_2 m_3}\mathcal{G}^{l_1l_2 H_2}_{m_1 m_2 -N_2} \mathcal{G}^{l_1 l_3 H_3}_{-m_1 m_3 -N_3}
\nonumber \\
&=f_{H_1l_2l_3}f_{H_2l_1l_2}f_{H_3l_1l_3}
(-1)^{H_1+l_2+l_3 + H_3+l_1+l_3}
\sum_{m_1m_2m_3}(-1)^{m_1+m_2+m_3}
\,K^{H_1 l_2 l_3}_{-N_1 m_2 -m_3}
K^{l_1 H_3 l_3}_{-m_1 -N_3 m_3}
K^{l_1 l_2 H_2}_{m_1 -m_2 -N_2}
\nonumber \\
&=f_{H_1l_2l_3}f_{H_2l_1l_2}f_{H_3l_1l_3} (-1)^{l_1+l_2+l_3} K^{H_1H_2H_3}_{-N_1-N_2-N_3}\begin{Bmatrix}H_{1} &H_{2} & H_{3}\\
l_{1} &l_{3} &l_{2}\end{Bmatrix}\,,
\end{align}
where we also used that $l_1+l_2+H_2$ must be even. Then,  after expressing all Gaunt factors in terms of Wigner coefficients, Eq.~\ref{eq:B3} becomes
\ba
\mathcal{C}^{L_1}_{M_1}{}^{L_2}_{M_2}{}^{\ell}_{m}{}^{,L_3 L_4}_{,l_1 l_2 l_3}
&=i^{-L_1 -L_2 +L_3+L_4}\sum_{L,H_1,H_2,H_3}f_{H_1l_2l_3}f_{H_2l_1l_2}f_{H_3l_1l_3} (-1)^{l_1+l_2+l_3}
\begin{Bmatrix}H_{1} &H_{2} & H_{3}\\
l_{1} &l_{3} &l_{2}\end{Bmatrix}(-1)^{m}
\nonumber \\
&\times
f_{L_3L_4L}f_{L\ell H_1}f_{H_2 L_3 L_1}f_{H_3 L_4 L_2}
\sum_{M_3,M_4,M}
{K}^{L_3L_4L}_{M_3M_4M}{K}^{L\ell H_1}_{-M m -N_1}  {K}^{H_2 L_3 L_1}_{N_2 M_3 M_1}{K}^{H_3 L_4 L_2}_{N_3 M_4 M_2} K^{H_1H_2H_3}_{-N_1-N_2-N_3} \nonumber \\
&=i^{-L_1 -L_2 +L_3+L_4}{K}^{L_1 L_2 \ell}_{M_1 M_2 -m}(-1)^{m}(-1)^{l_1+l_2+l_3} \nonumber \\
&\times 
\sum_{L,H_1,H_2,H_3}f_{H_1l_2l_3}f_{H_2l_1l_2}f_{H_3l_1l_3} f_{L_3L_4L}f_{L\ell H_1}f_{H_2 L_3 L_1}f_{H_3 L_4 L_2}(-1)^{H_1+H_2+H_3} \nonumber \\ 
&\times 
\begin{Bmatrix}H_{1} &H_{2} & H_{3}\\
l_{1} &l_{3} &l_{2}\end{Bmatrix}
\begin{Bmatrix}H_2&L_3&L_1\\
H_3&L_4&L_2\\
H_1&L&\ell\end{Bmatrix} 
\label{eq:coeff_C_simplified} \,,
\ea
where we used Eq.~\ref{eq:9j} to obtain the last line. 

Finally, substituting Eq.~\ref{eq:coeff_C_simplified} in Eq.~\ref{eq:wlll} and collecting constant factors, we obtain:
\ba
\mathcal{V}^{ \ell L_1 L_2}_{m M_1 M_2}(k,k_1,k_2) \equiv (32\pi)^{\frac{3}{2}}kk_1k_2
\begin{pmatrix}L_{1}  &L_{2} & \ell\\
M_{1} & M_{2} &-m\end{pmatrix}
(-1)^{m}\sum_{l_1l_2l_3L_3L_4}g^{L_1 L_2 \ell}_{l_1l_2l_3L_3L_4} J^{\ell L_3L_4}_{l_1l_2l_3}(k,k_1,k_2) 
\label{eq:Vlll_def} \,,
\ea
where we have defined
\begin{align}
J^{\ell L_3L_4}_{l_1l_2l_3}(k,k_1,k_2)\equiv
\int\dd r\,r^2
\,j_\ell(kr)
\,j_{L_3}(k_1r)
\,j_{L_4}(k_2r)
\,W(r) \,D^2(r)
\,Z_{l_1 l_2 l_3}(k_1,k_2,r) \,,
\label{eq:J_integral}
\end{align}
and
\begin{align}
g^{L_1 L_2 \ell}_{l_1l_2l_3L_3L_4} \equiv 
\frac{(-1)^{l_1+l_2+l_3}}{\(2l_1 + 1\)\(2l_2 + 1\)\(2l_3 + 1\)}  
i^{-L_1 -L_2 +L_3+L_4}\sum_{L,H_1,H_2,H_3}f_{H_1l_2l_3}f_{H_2l_1l_2}f_{H_3l_1l_3} f_{L_3L_4L}f_{L\ell H_1}f_{H_2 L_3 L_1}f_{H_3 L_4 L_2}\nonumber \\\times\begin{Bmatrix}H_{1} &H_{2} & H_{3}\\
l_{1} &l_{3} &l_{2}\end{Bmatrix}\begin{Bmatrix}H_2&L_3&L_1\\
H_3&L_4&L_2\\
H_1&L&\ell\end{Bmatrix} \,,
\label{eq:gcoeffdef}
\end{align}
which is real. There are $9$ triplets $(l_1,l_2,l_3)$ in the Legendre decomposition of the kernel $Z_2$, excluding contributions from the kernel $G_2$ and from terms proportional to $f_\mathrm{NL}$. As a result, for fixed $L_1,L_2,\ell$ there are at most $49$ terms in the sum of Eq.~\ref{eq:Vlll_def}. Finally, angle-averaging with Eq.~\ref{eq:wll_angle_averaged}, we obtain:
\ba
\mathcal{V}^{\ell L_1 L_2}(k,k_1,k_2) \equiv (32\pi)^{\frac{3}{2}}kk_1k_2
\sum_{l_1l_2l_3L_3L_4}g^{L_1 L_2 \ell}_{l_1l_2l_3L_3L_4} J^{\ell L_3L_4}_{l_1l_2l_3}(k,k_1,k_2) \,.
\label{eq:wlll_averaged}
\ea

\subsection{Contribution from the $f_\mathrm{NL}$ and $G_2$ terms}
\label{app:velocity_div_kernel}

Here we address the terms in the kernel $Z_2$ (Eq.~\ref{eq:z2_with_fnl}) that were left out when decomposing $Z_2$ as polynomial in $\khat_1\cdot\khat_2$, $\khat_1\cdot\rhat$, and $\khat_2\cdot\rhat$  (Eq.~\ref{eq:z2_decomposed}).  These terms are 
\ba
f_{\mathrm{NL}} \frac{\alpha(k,r)}{\alpha\left(k_{1},r\right) \alpha\left(k_{2},r\right)}\left(b^\mathrm{E}_{10} + f(r)\mu^2\right)
\label{eq:prob1} \,,
\ea
and
\ba
f(r) \mu^{2}G_{2}\left(\vk_{1}, \vk_{2}\right)
\label{eq:prob2} \,.
\ea
To account for them, it will be advantageous to first write the SFB bispectrum in an alternate form (Eq.~\ref{eq:SFB_bispectrum_radial}), which makes clear the relation of the observed bispectrum to the bispectrum of an isotropic and homogeneous Universe.

\subsubsection{Relation between position-dependent bispectrum and SFB bispectrum}

The SFB bispectrum is obtained from the position-dependent Fourier-space bispectrum by first transforming Eq.~\ref{eq:3point_bispectrum_fourier_final} to configuration space using Eq.~\ref{eq:observed_galaxy_density_general}, and then transforming into SFB space using Eq.~\ref{eq:sfb_fourier_pair_b}. We get
\ba
\<\delta^{\obs}_{g,l_1m_1}(k_1)\delta^{\obs}_{g,l_2m_2}(k_2)\delta^{\obs}_{g,l_3m_3}(k_3)\>=\bigg(\frac{2}{\pi}\bigg)^{\frac{3}{2}}k_1k_2k_3\int \bigg(\prod_i \frac{1}{(2\pi)^3}r_i^2\dd r_i q_i^2\dd q_i \dd^2\rhat_i\dd^2\qhat_iW(\vr_i)e^{i\vq_i\cdot\vr_i}j_{l_i}(k_ir_i)Y^*_{l_i,m_i}(\rhat_i)\bigg) \nonumber
\\ \times B_s(\vq_1,\vq_2,\vq_3,\vr_1,\vr_2,\vr_3) (2\pi)^3\delta_D(\vq_1+\vq_2+\vq_3) \,.
\label{eq:SFB_bispectrum_observed_corrected}
\ea

In the absence of RSD, linear growth, galaxy bias, and window, the observed SFB bispectrum (Eq.~\ref{eq:SFB_bispectrum_observed_corrected}) reduces to the SFB bispectrum in an isotropic and homogeneous Universe (Eq.~\ref{eq:sfb_3point_expanded}), using Eq.~\ref{eq:jljlDelta}. 
Unlike in the isotropic and homogeneous case, however, fixing the lengths $q_i$ and imposing $\vq_1+\vq_2+\vq_3=0$ fixes the angles $\qhat_i \cdot \qhat_j$ but does not determine $B_s(\vq_1,\vq_2,\vq_3,\vr_1,\vr_2,\vr_3)$, which depends on the $9$ angles $\mu_{ij}\equiv\qhat_i \cdot \rhat_j$ to the three lines of sight $\rhat_j$.

Assuming a radial window $W(\vr)=W(r)$, the angle-averaged bispectrum is given by
\ba
B^{\text{SFB, obs}}_{l_1l_2l_3}(k_1,k_2,k_3)=\frac{1}{(2\pi)^6}\bigg(\frac{2}{\pi}\bigg)^{\frac{3}{2}}k_1k_2k_3\int \bigg(\prod_i r_i^2\dd r_i q_i^2\dd q_i W(r_i)j_{l_i}(k_ir_i)\bigg)\mathcal{I}_{l_1l_2l_3}^{\text{ang.}}(q_1,q_2,q_3,r_1,r_2,r_3) \,,
\label{eq:SFB_bispectrum_radial}
\ea
where we have defined the angle-averaged angular integral
\ba
\mathcal{I}_{l_1l_2l_3}^{\text{ang.}}(q_1,q_2,q_3,r_1,r_2,r_3)
&\equiv
\sum_{m_1 m_2 m_3}
\begin{pmatrix}
l_1 & l_2 &  l_3 \\
  m_1 & m_2 & m_3
\end{pmatrix}
\int \bigg(\prod_i \dd^2\rhat_i\dd^2\qhat_ie^{i\vq_i\cdot\vr_i}Y^*_{l_i,m_i}(\rhat_i)\bigg)
\vs&\quad\times
B_s(\vq_1,\vq_2,\vq_3,\vr_1,\vr_2,\vr_3) \delta_D(\vq_1+\vq_2+\vq_3) \,.
\label{eq:SFB_bispectrum_angular}
\ea
{The integral $\mathcal{I}^\mathrm{ang.}_{l_1 l_2 l_3}$ is closely related to the TSH bispectrum, and would be the same if we were to integrate over the $q_i$. However, we use the above definition for clarity later in this appendix.

\subsubsection{$G_2$ contribution}

The contribution from the velocity kernel $G_2$ in the bispectrum $B_s$ (Eq.~\ref{eq:Bs}) is given by
\ba
B_s \supset D_1D_2D_3^2\left(2P(q_1)P(q_2)\right)\bigg[b_1+f_1\mu_{11}^2 \bigg]\bigg[b_2+f_2\mu_{22}^2 \bigg]\bigg[f_3 G_2 (\vq_1,\vq_2)\mu_{33}^2\bigg] +\mathrm{2\ cyc. \ perm.}\,,
\label{eq:G2_contrbution}
\ea
where we  write for brevity in this section $b_i=b^\mathrm{E}_{10}(r_i)+b^\mathrm{E}_{01}(r_i)/\alpha(k_i,r_i)$, $f_i=f(r_i)$, $D_i=D(r_i)$. 

We perform the integrals over $\rhat_i$ in Eq.~\ref{eq:SFB_bispectrum_angular} using the  identity Eq.~\ref{eq:spherical_transform_g1}. As $G_2(\vq_1,\vq_2)$ is rotationally invariant, when $\vq_1+\vq_2+\vq_3=0$ we may write it as  $ G_2 (\vq_1,\vq_2)=G_2(q_1,q_2,q_3)$, evaluated with $\qhat_j \cdot \qhat_k = \vartheta(q_j,q_k,q_l)$.  The $G_2$ contribution to Eq.~\ref{eq:SFB_bispectrum_angular} is thus
\ba
\mathcal{I}_{l_1l_2l_3}^{\text{ang.}}
&\supset
\sum_{m_1 m_2 m_3}
\begin{pmatrix}
  l_1 & l_2 &  l_3 \\
  m_1 & m_2 & m_3
\end{pmatrix}
\int \left(\prod_j \dd^2\qhat_j\right) \delta_D(\vq_1+\vq_2+\vq_3)
\bigg\{
\vs&\quad\times
\bigg[2\prod_{i=1,2} D_i P(q_i)4\pi i^{l_i} Y^*_{l_i,m_i}(\qhat_i)\bigg(b_ij_{l_i}(q_ir_i)-f_ij_{l_i}''(q_ir_i)\bigg)\bigg]
\nonumber\\&\quad\times
D_3^2\bigg[-4\pi i^{l_3} Y^*_{l_3,m_3}(\qhat_3)f_3 G_2 (\vq_1,\vq_2)j_{l_3}''(q_3r_3)\bigg] +\mathrm{2\ cyc. \ perm.}\bigg\} \nonumber \\
&=(4\pi)^3i^{l_1+l_2+l_3} 
\bigg[2\prod_{i=1,2} P(q_i)
D_i\bigg(b_ij_{l_i}(q_ir_i)-f_ij_{l_i}''(q_ir_i)\bigg)\bigg]
D_3^2\bigg(-f_3 G_2 (q_1,q_2,q_3)j_{l_3}''(q_3r_3)\bigg) \nonumber \\
&\quad\times
\sum_{m_1 m_2 m_3}
\begin{pmatrix}
l_1 & l_2 &  l_3 \\
  m_1 & m_2 & m_3
\end{pmatrix}
\int d^2\qhat_1d^2 \qhat_2d^2\qhat_3Y^*_{l_1,m_1}(\qhat_1)Y^*_{l_2,m_2}(\qhat_2)Y^*_{l_3,m_3}(\qhat_3)\delta_D(\vq_1+\vq_2+\vq_3) +\mathrm{2\ cyc. \ perm.}
\label{eq:Iii} \,.
\ea
We recognize the integral in the last line of Eq.~\ref{eq:Iii} as the integral defined in Eq.~\ref{eq:gauntlike_identity_app}, which is proportional to a Wigner-$3j$ symbol. Thus, the sum simplifies with the identity Eq.~\ref{eq:gaunt_orthogonality}, and we can express the last line as the integral $I_{l_1l_2l_3}(q_1,q_2,q_3)$ of Eq.~\ref{eq:Il1l2l3}. We obtain the $G_2$ contribution to the SFB bispectrum by inserting Eq.~\ref{eq:Iii} into Eq.~\ref{eq:SFB_bispectrum_radial}. The integrals over $r_1$ and $r_2$ can be written compactly in terms of the kernels $\mathcal{W}_l(k,q)$ (Eq.~\ref{eq:W_lkq_first}) as
\ba
&B^{\text{SFB}, G_2}_{l_1l_2l_3}(k_1,k_2,k_3)
\vs
&=
\frac{1}{(2\pi)^6}\bigg(\frac{2}{\pi}\bigg)^{\frac{3}{2}}k_1k_2k_3(4\pi)^3i^{l_1+l_2+l_3}\int \bigg(\prod_i dq_i q_i^2\bigg)2P(q_1)P(q_2)\bigg(\frac{\pi}{2k_1q_1}\mathcal{W}_{l_1}(k_1,q_1)\bigg)\bigg(\frac{\pi}{2k_2q_2}\mathcal{W}_{l_2}(k_2,q_2)\bigg)\nonumber \\
&\quad\times G_2(q_1,q_2,q_3)\bigg(\int dr_3 r_3^2W(r_3)j_{l_3}(k_3r_3)D^2(r_3)(-f(r_3)j''_{l_3}(q_3r_3))\bigg)I_{l_1l_2l_3}(q_1,q_2,q_3) + \mathrm{2\ cyc.\  perm.}\,.
\label{eq:BSFBG2}
\ea

\subsubsection{Total contribution from $f_\mathrm{NL}$ and $G_2$ terms}

The contribution from the $f_\mathrm{NL}$ term  Eq.~\ref{eq:prob1} is analogous to Eq.~\refeq{eq:BSFBG2}. Noting that we may factorize $\alpha(k,r)=\gamma(k)D(r)$ with
\ba
\gamma(k) \equiv \frac{2k^2c^2T(k)}{3\Omega_mH_0^2} \,,
\ea
we may define, in analogy to the derivation in \ref{sec:main_derivation}, the kernels \footnote{Note that $\mathcal{W}_l(k,q)$ (Eq.~\ref{eq:W_lkq_first}) differs from Eq.~\ref{eq:wlfnl}  in that the scale-dependent bias $b(r,q)$ is replaced by $b^\mathrm{E}_{10}(r)$.} 
\ba
\mathcal{W}^{G_2}_\ell(k,q)&\equiv \frac{2kq}{\pi}\int dr \, r^2W(r)j_{\ell}(kr)D^2(r)\bigg(-f(r)j''_{\ell}(qr)\bigg) \\
\mathcal{W}^{f_\mathrm{NL}}_\ell(k,q)&\equiv \frac{2kq}{\pi}\int dr \, r^2W(r)j_{\ell}(kr)D(r)\bigg(b^\mathrm{E}_{10}(r)j_{\ell}(qr)-f(r)j''_{\ell}(qr)\bigg) \label{eq:wlfnl}\\
\mathcal{V}_{f_{\mathrm{NL},G_2}}^{l_3l_1l_2}(k_3,q_1,q_2)&\equiv \frac{1}{(2\pi)^{\frac{3}{2}}}i^{l_1+l_2+l_3}\int dq_3 (q_1q_2q_3)I_{l_1l_2l_3}(q_1,q_2,q_3) \nonumber \\
& \times \bigg[G_2(q_1,q_2,q_3)\mathcal{W}^{G_2}_{l_3}(k_3,q_3) +\Bigg(f_{\mathrm{NL}} \frac{\gamma(q_3)}{\gamma\left(q_{1}\right) \gamma\left(q_{2}\right)}\bigg)\mathcal{W}_{l_3}^{f_\mathrm{NL}}(k_3,q_3)\bigg] \,.
\label{eq:wlllG2}
\ea  
Finally, after reordering the cyclic permutations to match the ordering in the main text, the combined contribution to the bispectrum signal from the $f_\mathrm{NL}$ and $G_2$ terms is
\ba
B^{\text{SFB,}\,f_{\mathrm{NL}},G_2}_{l_1l_2l_3}(k_1,k_2,k_3)=
2\int\dd q_2
\mathcal{W}_{l_2}(k_2,q_2)
\,P(q_2)
\int\dd q_3
\,\mathcal{W}_{l_3}(k_3,q_3)
\,P(q_3) \ \mathcal{V}_{f_\mathrm{NL}, G_2}^{l_1l_2l_3}(k_1,q_2,q_3)+ \mathrm{2\ cyc.\  perm.}
\label{eq:sfb_G2_appendix}
\ea
The advantage of expressing Eq.~\ref{eq:wlllG2} in the above form is that the integral $I_{l_1l_2l_3}$ can be rapidly (pre)computed without numerical integration via the identity Eq.~\ref{eq:Il1l2l3}.

\clearpage
\end{document}